\documentclass{acm_proc_article-sp}
\usepackage{amsfonts,times,epsfig,latexsym}
\usepackage[tight,footnotesize]{subfigure}
\usepackage{algorithm}
\usepackage{algorithmic}

\makeatletter
\newcommand{\figcaption}{\def\@captype{figure}\caption}
\newcommand{\tabcaption}{\def\@captype{table}\caption}
\makeatother

\newtheorem{defn}{\textsc{Definition}}

\begin{document}

\title{HRank: A Path based Ranking Framework in Heterogeneous Information Network}
%\subtitle{[Extended Abstract]}

\numberofauthors{6} %  in this sample file, there are a *total*
\author{
%\alignauthor
Yitong Li,
%       \affaddr{Beijing University of Posts and Telecommunications}\\
%       \affaddr{Beijing, China}\\
%       \email{yitongglee@gmail.com}
%\alignauthor
Chuan Shi,
%       \affaddr{Beijing University of Posts and Telecommunications}\\
%       \affaddr{Beijing, China}\\
%       \email{shichuan@bupt.edu.cn}
%\and
%\alignauthor
Philip S. Yu, and
%       \affaddr{University of Illinois at Chicago}\\
%       \affaddr{IL, USA}\\
%       \email{psyu@cs.uic.edu}
%\alignauthor
Qing Chen
%       \affaddr{China Mobile Communications Corporation}\\
%       \affaddr{Beijing, China}\\
%       \email{chenqing@chinamobile.com}
}

\toappear{Y.T. Li, C. Shi are with Beijing University of Posts and Telecommunications, Beijing, China. \\
E-mail: yitongglee@gmail.com, shichuan@bupt.edu.cn.\\
P.S. Yu is with University of Illinois at Chicago, IL, USA. \\
E-mail: psyu@uic.edu.\\
Q. Chen is with China Mobile Communications Corporation, Beijing, China. E-mail: chenqing@chinamobile.com.}

%\toappear{Permission to make digital or hard copies of all or part
%of this work for personal or classroom use is granted without fee
%provided that copies are not made or distributed for profit or
%commercial advantage and that copies bear this notice and the full
%citation on the first page. To copy otherwise, to republish, to post
%on servers or to redistribute to lists, requires prior specific
%permission and/or a fee. \par {\confname EDBT 2012}, March 26--30,
%2012, Berlin, Germany.\par Copyright 2012 ACM
%978-1-4503-0790-1/12/03 ...\$10.00}

\maketitle
\begin{abstract}
Recently, there is a surge of interests on heterogeneous information network analysis. As a newly emerging network model, heterogeneous information networks have many unique features (e.g., complex structure and rich semantics) and a number of interesting data mining tasks have been exploited in this kind of networks, such as similarity measure, clustering, and classification. Although evaluating the importance of objects has been well studied in homogeneous networks, it is not yet exploited in heterogeneous networks. In this paper, we study the ranking problem in heterogeneous networks and propose the HRank framework to evaluate the importance of multiple types of objects and meta paths. Since the importance of objects depends upon the meta paths in heterogeneous networks, HRank develops a path based random walk process. Moreover, a constrained meta path is proposed to subtly capture the rich semantics in heterogeneous networks. Furthermore, HRank can simultaneously determine the importance of objects and meta paths through applying the tensor analysis. Extensive experiments on three real datasets show that HRank can effectively evaluate the importance of objects and paths together. Moreover, the constrained meta path shows its potential on mining subtle semantics by obtaining more accurate ranking results.
\end{abstract}



 % NOT required for Proceedings

\section{Introduction}
It is an important research problem to evaluate object importance or popularity, which can be used in many data mining tasks. Many methods have been developed to evaluate object importance, such as PageRank \cite{PBMW98}, HITS \cite{K99}, and SimRank \cite{JW02}. In these literatures, objects ranking is done in a homogeneous network in which objects or relations are the same. For example, both PageRank and HITS rank the web page in WWW.

However, in many real network data, there are many different types of objects and relations, which can be organized as heterogeneous network. Formally, Heterogeneous Information Networks (HIN) are the logical networks involving multiple types of objects as well as multiple types of links denoting different relations \cite{Han09}. For example, the movies recommendation data include multiple types of objects: movies, actors, and directors and their relations \cite{SKYX12}. It is clear that heterogeneous information networks are ubiquitous and form a critical component of modern information infrastructure \cite{Han09}. Recently, many data mining tasks have been exploited in this kind of networks, such as similarity measure \cite{SHYYW11,SKYX12}, clustering \cite{SYH09, SHZYCW09}, and classification \cite{JSDHG10}, among which ranking is an important but not yet exploited task.

%\vspace{-10pt}
\begin{figure}[htbp]
\centering
\small
\subfigure[Heterogeneous network]{\label{fig:fft:a}
\begin{minipage}[t]{0.27\textwidth}

  \includegraphics[width=4.5cm]{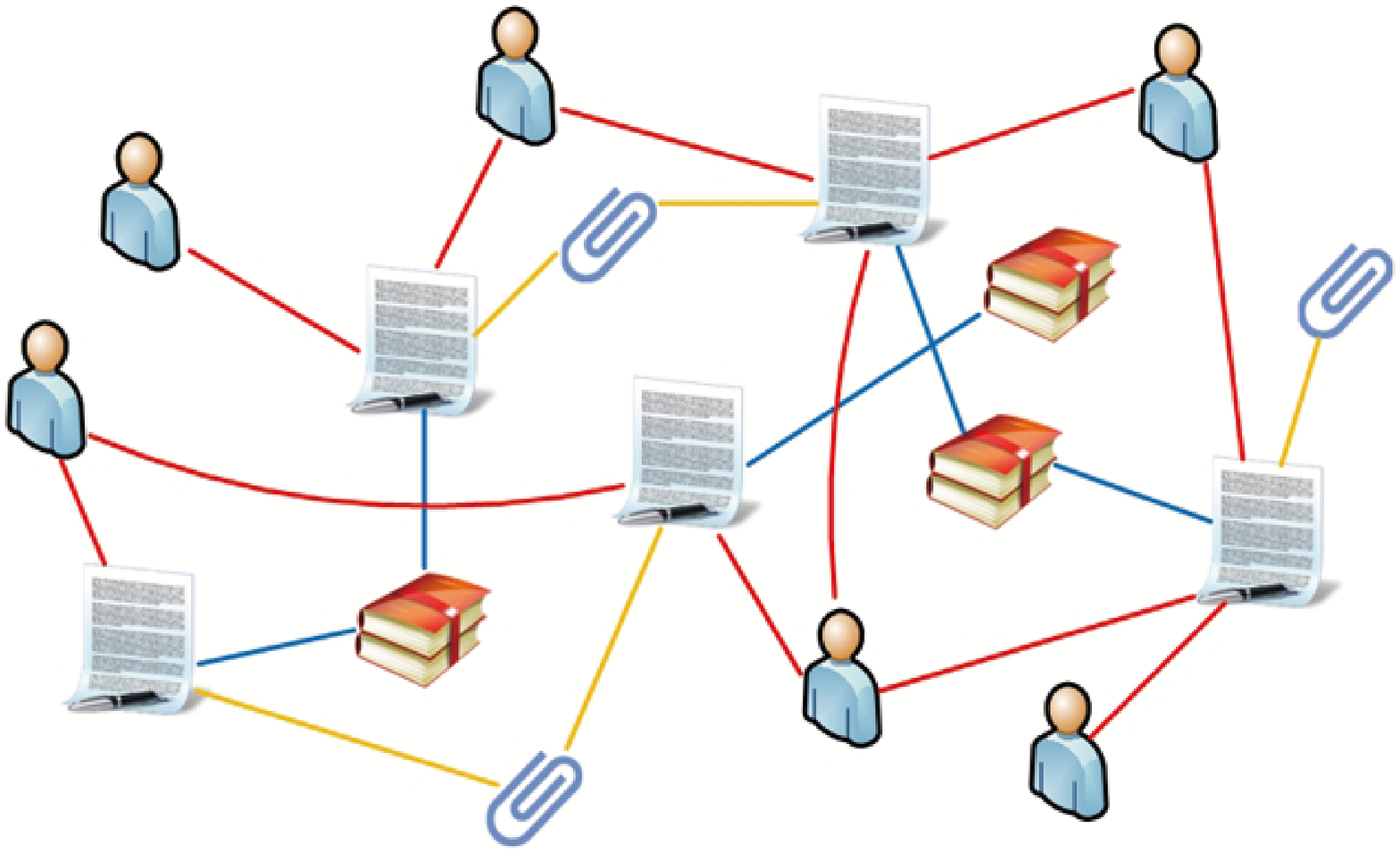}
\end{minipage}%
}%
\subfigure[Network schema]{
\begin{minipage}[t]{0.19\textwidth}

  \includegraphics[width=3.0cm]{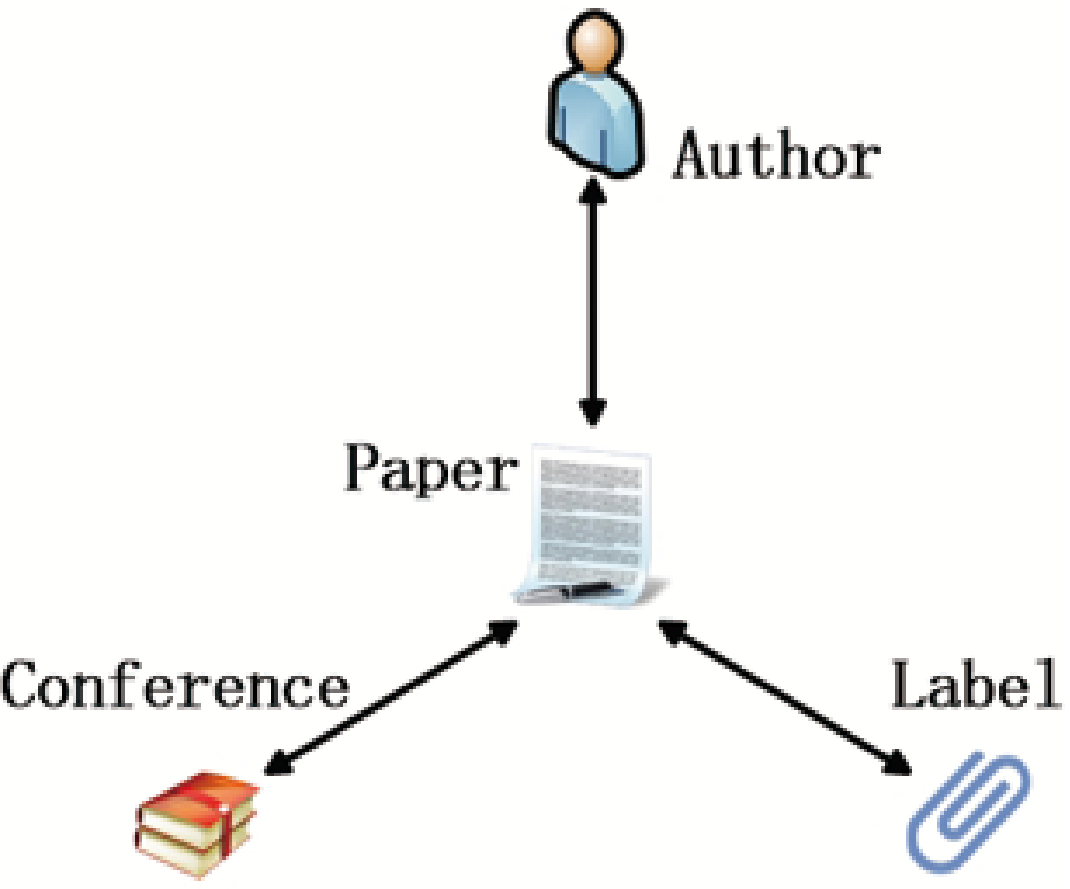}
\end{minipage}
} %\vspace{-10pt}
\caption{\small A heterogeneous information network example on bibliographic data. (a) shows heterogeneous objects and their relations. (b) shows the network schema.}\label{fig:fft}
\end{figure}
%\vspace{-10pt}

Figure 1(a) shows a HIN example in bibliographic data and Figure 1(b) illustrates its network schema. In this example, it contains objects from four types of objects: papers ($P$), authors ($A$), labels ($L$, categories of papers) and conferences ($C$). There are links connecting different types of objects. The link types are defined by the relations between two object types. For example, links exist between authors and papers denoting the writing or written-by relations, between conferences and papers denoting the publishing or published-in relations. In this network, several interesting, yet seldom exploited, ranking problems can be proposed.

\begin{itemize}
\item
One may be interested in the importance of one type of objects, and ask the following questions:\\
\small{
\emph{Q. 1.1 Who are the most influential authors?\\
Q. 1.2 Who are the most influential authors in data mining field? \\}}
\item \normalsize
As we know, some object types have an effect on each other. For example, influential authors usually publish papers in reputable conferences. So one may pay attention to the importance of multiple types of objects simultaneously, and ask the following questions:\\
\small{
\emph{Q. 2.1 Which are the most influential authors and reputable conferences?\\
Q. 2.2 Which are the most influential authors and reputable conferences in data mining field?\\}}
\item \normalsize
Furthermore, one may wonder which factor mostly affects the importance of objects, since the importance of objects is affected by many factors. So he may ask the questions like this:\\
\small{
\emph{Q. 3 Who are the most influential authors and which factor makes the author most influential?\\}}
\end{itemize}

Although the ranking problem in homogeneous networks has been well studied, the above ranking problems are unique in HIN (especially \emph{Q. 2} and \emph{Q. 3}), which are seldom studied until now. Since there are multiple types of objects in HIN, it is possible to analyze the importance of multiple types of objects (i.e., \emph{Q. 2}) as well as affecting factors (i.e., \emph{Q. 3}) together. However, the ranking analysis in HIN also faces the following research challenges.
\begin{itemize}
\item
 There are different types of objects and links in HIN. If we simply treat all objects equally and apply the random walk as PageRank does in homogeneous network, the ranking result will mix different types of objects together. It is difficult to distinguish the importance of a specific type of objects. Moreover, the result has obvious bias on object types which have many links.
\item
Different types of objects and links in heterogeneous
networks carry different semantic meanings. The random walk along different meta paths has different semantics, which may lead to different ranking results. Here the meta path \cite{SHYYW11} means a sequence of relations between object types. For example, the importance of authors should be different based on the relations of co-author (i.e., the meta path ``\emph{Author-Paper-Author}'') and co-conference (i.e., the meta path ``\emph{Author-Paper-Conference-Paper-Author}''). So a desirable ranking method in HIN should be path-dependent, which can capture the semantics under specific meta paths and return different values based on specific meta paths.
\end{itemize}

In this paper, we study the ranking problem in HIN and propose a novel ranking framework, HRank, to evaluate the importance of multiple types of objects and meta paths in HIN. For \emph{Q. 1} and \emph{Q. 2}, a path based random walk model is proposed to evaluate the importance of single or multiple types of objects. That is, the random walk process in HIN should follow the meta path designated beforehand. The different meta paths connecting two types (same or different) of objects have different semantics and transitive probability, and thus lead to different random walk processes and ranking results. Although meta path has been widely used to capture the semantics in HIN \cite{SKYX12, SHYYW11}, it coarsely depicts object relations. By employing the meta path, we can answer the \emph{Q. 1.1} and \emph{Q. 2.1}, but cannot answer the \emph{Q. 1.2} and \emph{Q. 2.2}. For example, ``\emph{Author-Paper-Author}'' describes the collaboration relation among authors. However, it cannot depict the fact that Philip S. Yu and Jiawei Han have many collaborations in data mining field but they seldom collaborate in information retrieval field. In order to overcome the shortcoming existing in meta path, we propose the \emph{constrained meta path} concept, which can effectively describe this kind of subtle semantics. The constrained meta path sets constraint conditions on meta path. Through adopting the constrained meta path, we can answer the \emph{Q. 1.2} and \emph{Q. 2.2}.

In HIN, there are many constrained meta paths connecting two types of objects. Based on different paths, the objects have different ranking values. The comprehensive importance of objects should consider all kinds of factors (the factors can be embodied by constrained meta paths). Moreover, these constrained meta paths have different contribution to the importance of objects. For example, although Jiawei Han and W. Bruce Croft both are influential authors in computer science, but the achievements on data mining and information retrieval fields contribute to their reputation, respectively. In order to evaluate the importance of objects and meta paths simultaneously (i.e., answer \emph{Q. 3}), we further propose a co-ranking method which organizes the relation matrices of objects on different constrained meta paths as a tensor. A random walk process is designed on this tensor to co-rank importance of objects and paths simultaneously. That is, random walkers surf in the tensor, where the stationary visiting probability of objects and meta paths is considered as the HRank score of objects and paths. In addition, in order to fasten the matrix multiplication process in HRank, we design three fast computation strategies whose effectiveness have been validated by experiments.

In all, this paper has the following three contributions.
\begin{itemize}
\item
We propose the constrained meta path concept to describe the subtle semantic relation in HIN. Compared to the meta path, the constrained meta path can depict object relation with finer granular through setting constraint condition on meta path.
\item
We propose a path-based ranking method to evaluate the importance of same or different types of objects in HIN by setting constrained meta path. Furthermore, we propose a co-ranking method to simultaneously evaluate the importance of objects and paths. The method not only can comprehensively evaluate the importance of objects by considering all constrained meta paths, but also can rank the contribution of different constrained meta paths.
\item
Extensive experiments evaluate the proposed method. We not only validate that the objects have different importance based on different constrained meta paths, but also show that the ranking results of constrained meta paths more comply with our common sense. In addition, the experimental results illustrate that the proposed method can accurately identify the importance of objects and the corresponding paths.
\end{itemize}

The rest of the paper is organized as follows. In Section
2, we summarize and compare the related work. In Section 3, we describe notations in this paper and some preliminary knowledge. In Section 4, we present the proposed method and the fast computation strategies are introduced in Section 5. Extensive experiments are done to validate the proposed method in Section 6. Finally,
Section 7 concludes this paper.

\section{Related work}
Ranking is an important data mining task in network analysis. Many ranking methods have been proposed. For example, PageRank \cite{PBMW98} evaluates the importance of objects through a random walk process; HITS \cite{K99} ranks objects using the authority and hub scores; SimRank \cite{JW02} evaluates the similarity of two objects by their neighbors' similarities. The recently proposed RoleSim measures the role similarity between any two nodes from network structure \cite{JLH11}. These approaches only consider the same type of objects in homogeneous networks, so they cannot be applied in heterogeneous networks. Some researches have begun to pay attention to the co-ranking on multiple types of objects. For example, Zhou et al. \cite{ZOZG07} co-rank authors and their publications by coupling two random walk processes, and the co-HITS \cite{DLK09} incorporates the bipartite graph with the content information and the constraints of relevance. Although these methods can rank different types of objects existing in HIN, they are restricted to bipartite graphs. Recently, MultiRank \cite{NLY11} determines the importance of both objects and relations simultaneously for multi-relational data, and HAR \cite{LNY12} is proposed to determine hub and authority scores of objects and relevance scores of relations in multi-relational data for query search. These two methods focus on same type of objects with multi-relations, not multiple types of objects.

In recent years, there is a surge on the HIN analysis. Many data mining tasks have been exploited in HIN, such as similarity measure \cite{SKYX12, SHYYW11}, clustering \cite{SHZYCW09, SYH09, SNHYYY12}, classification \cite{KYDW}. As an unique feature of HIN, the links connecting different types of objects contain semantics. So the meta path \cite{SHYYW11}, connecting object types via a sequence of relations, has been widely used to capture the relation semantics. Sun et al. \cite{SHYYW11} put forward the concept of meta path to describe the rich semantic relations, and studied similarity search on symmetric meta paths. As an extension of Sun's work, Yu et al. \cite{YSNMH12} use a meta-path-based ranking model ensemble to represent semantic meanings for similarity queries. HeteSim is also proposed by Shi et al. \cite{SKYX12} to measure the relevance scores of heterogeneous objects in HIN. PathSelClus \cite{SNHYYY12} integrates meta path selection with user-guided clustering to cluster objects in networks. Kong et al.\cite{KYDW} develop a HCC solution to assign labels to a group of instances that are interconnected through different meta paths. Although the meta path may convey semantic information in HIN, it is too coarse to capture the subtle semantics in some applications.

\section{Preliminary}
In this section, we describe notations used in this paper and present some preliminary knowledge.

A heterogeneous information network is a special type of information network with the underneath data structure as a directed graph, which either contains multiple types of objects or multiple types of links.

\begin{defn}
\textbf{Information Network} \cite{SHYYW11}. Given a schema $S=(\mathcal{A},\mathcal{R})$ which consists of a set of entity types $\mathcal{A} = \{A\}$ and a set of relations $\mathcal{R} = \{R\}$, an information network is defined as a directed graph $G = (V,E)$ with an object type mapping function $\varphi : V \rightarrow \mathcal{A}$ and a link type mapping function $\psi : E \rightarrow \mathcal{R}$. Each object $v \in V$ belongs to one particular object type $\varphi (v) \in \mathcal{A}$, and each link $e \in E$ belongs to a particular relation $\psi (e) \in \mathcal{R}$. When the types of objects $|\mathcal{A}| > 1$ or the types of relations $|\mathcal{R}| > 1$, the network is called \textbf{heterogeneous information network}; otherwise, it is a \textbf{homogeneous information network}.
\end{defn}

In heterogeneous information networks, there are multiple object types and relation
types. We use the network schema to depict the
object types and the relations existing among object types. For a
relation $R$ existing from type $S$ to type $T$, denoted as
$S\overset{R}{\longrightarrow}T$, $S$ and $T$ are the \textbf{source
type} and \textbf{target type} of relation $R$, which is denoted as
$R.S$ and $R.T$, respectively. The inverse relation $\emph{R}^{-1}$
holds naturally for $T\overset{\emph{R}^{-1}}{\longrightarrow}S$.
Generally, $\emph{R}$ is not equal to $\emph{R}^{-1}$, unless
$\emph{R}$ is symmetric and these two types are the same. Figure 1(b) shows a network schema of bibliographic information network, which describes the object types and their relations in the HIN.

Different from homogeneous networks, two objects in a heterogeneous network can be connected via different paths and these paths have different meanings. As an example shown in Figure 1(b), authors can be connected via ``Author-Paper-Author'' ($APA$) path, ``Author-Paper-Conference-Paper-Author'' ($APCPA$) path and so on. These paths are called meta paths which can be defined as follows.

\begin{defn}
\textbf{Meta path} \cite{SHYYW11}. A meta path $\mathcal{P}$ is a path defined on a schema $S=(\mathcal{A},\mathcal{R})$, and is denoted in the form of $A_1 \xrightarrow{R_1} A_2 \xrightarrow{R_2} \ldots \xrightarrow{R_l} A_{l+1}$ (abbreviated as $A_1A_2\ldots A_{l+1}$), which defines a composite relation $R = R_1 \circ R_2 \circ \ldots \circ R_l$ between type $A_1$ and $A_{l+1}$, where $\circ$ denotes the composition operator on relations.
\end{defn}

It is obvious that semantics underneath these paths are different. The \emph{APA} path means authors collaborating on the same papers (i.e., co-author relation), while the \emph{APCPA} path means the authors' papers publishing on the same conferences (i.e., co-conference relation). Based on different meta paths, there are different relation networks, which may result in different importance of objects. For example, the importance of authors under \emph{APA} path has bias on the authors who write many papers having many authors, while the importance of authors under \emph{APCPA} path emphasizes the authors who publish many papers on those productive conferences. So the importance of objects depends on the meta path in the heterogeneous networks. As an effective semantic capturing method, meta path has been widely used in many data mining tasks in HIN, such as similarity measure \cite{SKYX12, SHYYW11}, clustering \cite{SNHYYY12}, and classification \cite{KYDW}. However, meta path fails to capture some subtle semantics. Taking Figure 1(b) as an example, the \emph{APA} cannot reveal the co-author relations in Data Mining (DM) field.

In order to overcome the shortcomings in meta path, we propose the concept of constrained meta path, defined as follows.
\begin{defn}
\textbf{Constrained meta path}. A constrained meta path is a meta path based on a certain constraint which is denoted as $\mathcal{CP} = \mathcal{P}|\mathcal{C}$. $\mathcal{P} = (A_{1}A_{2}\ldots A_{l})$ is a meta path, while $\mathcal{C}$ represents the constraint on the objects in the meta path.
\end{defn}
Note that the $\mathcal{C}$ can be one or multiple constraint conditions on objects.  Taking Figure 1(b) as an example, the constrained meta path $APA|P.L=``DM"$ represents the co-author relations of authors in data mining field through constraining the label of papers with DM. Similarly, the constrained meta path $APCPA|P.L=``DM"\&\&C=``CIKM"$ represents the co-author relations of authors in CIKM conference and the papers of authors are in data mining field. Obviously, compared to meta path, the constrained meta path conveys richer semantics by subdividing meta paths under distinct conditions. Particularly, when the length of meta path is 1 (i.e., a relation), the constrained meta path degrades to a \textbf{constrained relation}. In other words, the constrained relation confines condition on objects of the relation.

For a relation $A\overset{R}{\longrightarrow}B$, we can obtain its transition probability matrix as follows.

\begin{defn}
\textbf{Transition probability matrix}. $W_{AB}$ is an adjacent matrix
between type $A$ and $B$ on relation $A\overset{R}{\longrightarrow}B$. $U_{AB}$ is the normalized matrix of $W_{AB}$ along the row vector, which is the transition probability matrix of
$A\overset{R}{\longrightarrow}B$.
\end{defn}

Then we make some constraints on objects of the relation $A\overset{R}{\longrightarrow}B$ (i.e., constrained relation). We can have the following definition.

\begin{defn}
\textbf{Constrained transition probability matrix}. $W_{AB}$ is an adjacent matrix
between type $A$ and $B$ on relation $A\overset{R}{\longrightarrow}B$. Suppose there is a constraint $\mathcal{C}$ on object type A. The constrained transition probability matrix $U^{'}_{AB}$ of constrained relation $R|\mathcal{C}$ is $U^{'}_{AB}=M_{\mathcal{C}}U_{AB}$, where $M_{\mathcal{C}}$ is the constraint matrix generated by the constraint condition $\mathcal{C}$ on object type A.
\end{defn}

The constraint matrix $M_{\mathcal{C}}$ is usually a diagonal matrix whose dimension is the number of objects in object type \emph{A}. The element in the diagonal is 1 if the corresponding object satisfies the constraint, else the element in the diagonal is 0. Similarly, we can confine the constraint on the object type \emph{B} or both types. Note that the transition probability matrix is a special case of the constrained transition probability matrix, when we let the constraint matrix $M_{\mathcal{C}}$ be the identity matrix $I$.

Given a network $G=(V,E)$ following a network schema $S=(\mathcal{A},\mathcal{R})$, we can define the meta path based reachable probability matrix as follows.

\begin{defn}
\textbf{Meta path based reachable probability matrix}.
For a meta path $\mathcal{P}=(A_1A_2\cdots A_{l+1})$, the meta path based reachable probability matrix $PM$ is defined as $PM_{\mathcal{P}}=U_{A_1A_2}U_{A_2A_3}\\ \cdots U_{A_{l}A_{l+1}}$. $PM_{\mathcal{P}}(i,j)$ represents the probability of object $i\in A_1$ reaching object $j\in A_{l+1}$ under the path $\mathcal{P}$.
\end{defn}

Similarly, we have the following definition for constrained meta path.

\begin{defn}
\textbf{Constrained meta path based reachable probability matrix}.
For a constrained meta path $\mathcal{CP}=(A_1A_2\cdots A_{l+1}|\mathcal{C})$, the constrained meta path based reachable probability matrix is defined as $PM_{\mathcal{CP}}=U^{'}_{A_1A_2}U^{'}_{A_2A_3}\cdots U^{'}_{A_{l}A_{l+1}}$. $PM_{\mathcal{CP}}(i,j)$ represents the probability of object $i\in A_1$ reaching object $j\in A_{l+1}$ under the constrained meta path $\mathcal{P|\mathcal{C}}$.
\end{defn}

In fact, if there is no constraint on the objects of a relation $A_{i}\overset{R}{\longrightarrow}A_{i+1}$, $U^{'}_{A_iA_{i+1}}$ is equal to $U_{A_iA_{i+1}}$. If there is a constraint on the objects, we only consider the objects that satisfy the constraint. For simplicity, we use the reachable probability matrix and the $M_{\mathcal{P}}$ to represent the constrained meta path based reachable probability matrix in the following section.

\section{The HRank Method}
Since the importance of objects is related to the meta path designated by users, we propose the path based ranking method HRank in heterogeneous networks. In order to answer the three ranking problems proposed in Section 1, we design three versions of HRank, respectively.

%\vspace{-10pt}
\begin{figure}[htbp]
\centering
\small

\begin{minipage}[t]{0.18\textwidth}
  \includegraphics[width=3cm]{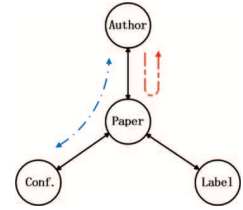}
\end{minipage}%
%\vspace{-5pt}
\caption{\small An example of the computation process of HRank. The blue and red broken line represents the process on the symmetric and asymmetric constrained meta path, respectively. }\label{fig:fft}
\end{figure}
%\vspace{-10pt}

\subsection{Ranking based on symmetric constrained meta paths}
In order to evaluate the importance of one type of objects (i.e., \emph{Q. 1}), we design the HRank-SY method based on symmetric constrained meta paths, since the constrained meta paths connecting one type of objects are usually symmetric, such as $APA|P.L=``DM"$.

For a symmetric constrained meta path $\mathcal{P} = (A_{1}A_{2}\ldots A_l|C)$ , $\mathcal{P}$ is equal to $\mathcal{P}^{-1}$ and $A_1$ and $A_l$ are the same. Similar to PageRank \cite{PBMW98}, the importance evaluation of object $A_1$ (i.e., $A_l$) can be considered as a random walk process in which random walkers wander from type $A_1$ to type $A_l$ along the path $\mathcal{P}$. The HRank value of object $A_1$ ($R(A_1|\mathcal{P})$) is the stable visiting probability of random walkers, which is defined as follows:
%\vspace{-5pt}
\begin{equation}
\small
R(A_1|\mathcal{P}) = \alpha R(A_1|\mathcal{P})M_{\mathcal{P}} + (1-\alpha)E
\end{equation} %\vspace{-3pt}
where $M_\mathcal{P}$ is the constrained meta path based reachable probability matrix as defined above. $E$ is the restart probability for convergence, which is set equally. $\alpha$ is the weight of restart, which can be set with 0.15 as PageRank does.  HRank-SY and PageRank both have the same idea that the importance of objects is decided by the visiting probability of random surfers. Different from PageRank, the random surfers in HRank-SY should wander along the constrained meta path to visit objects.

As shown in Figure 2, the red broken line illustrates an example of the process of calculating rank value, where the $\mathcal{CP}$ is $APA|P.L=``DM"$. The concrete calculating process is as follows.
\begin{equation}
\small
\begin{split}
R(Author|\mathcal{CP}) &= \alpha R(Author|\mathcal{CP}) M_{\mathcal{CP}}+ (1-\alpha)E\\
M_{\mathcal{CP}} &= U^{'}_{AP}U^{'}_{PA}=U_{AP}M_{P}M_{P}U_{PA}
\end{split}
\end{equation}
where $M_{P}$ is the constraint matrix on object type P (paper).

\subsection{Ranking based on asymmetric constrained meta paths}
For the question \emph{Q. 2}, we propose the HRank-AS method based on asymmetric constrained meta paths, since the paths connecting different types of objects are asymmetric. For an asymmetric constrained meta path $\mathcal{P} = (A_{1}A_{2}\ldots A_l|\mathcal{C})$, $\mathcal{P}$ is not equal to $\mathcal{P}^{-1}$. Note that $A_1$ and $A_l$ are either of the same or different types, such as $APC|P.L=``DM"$ and $PCPLP|C=``CIKM"$.

Similarly, HRank-AS is also based on a random walk process that random walkers wander between $A_1$ and $A_l$ along the path. The ranks of $A_1$ and $A_l$ can be seen as the visiting probability of walkers, which are defined as follows:
\begin{equation}
\small
\begin{split}
R(A_l|\mathcal{P}^{-1}) = \alpha R(A_1|\mathcal{P})M_{\mathcal{P}} +(1-\alpha)E_{A_l}
\\
R(A_1|\mathcal{P}) = \alpha R(A_l|\mathcal{P}^{-1})M_{\mathcal{P}^{-1}} +(1-\alpha)E_{A_1}
\end{split}
\end{equation}
where $M_\mathcal{P}$ and $M_{\mathcal{P}^{-1}}$ are the reachable probability matrix of path $\mathcal{P}$ and $\mathcal{P}^{-1}$. $E_{A_1}$ and $E_{A_l}$ are the restart probability of $A_1$ and $A_l$. Obviously, HRank-SY is the special case of HRank-AS. When the path $\mathcal{P}$ is symmetric, the Eq. 3 is the same with Eq. 1.

The blue broken line in Figure 2 illustrates an example which simultaneously evaluates the importance of authors and conferences. Here the $\mathcal{CP}$ is $APC|P.L=``DM"$. The concrete calculating process is as follows.
\begin{equation}
\small
\begin{split}
R(Conf.|\mathcal{CP}) &= \alpha R(Aut.|\mathcal{CP}) M_{\mathcal{CP}}+ (1-\alpha)E_{Conf.}\\
R(Aut.|\mathcal{CP}) &= \alpha R(Conf.|\mathcal{CP}) M_{\mathcal{CP}^{-1}}+ (1-\alpha)E_{Aut.}\\
M_{\mathcal{CP}} &= U^{'}_{AP}U^{'}_{PC}=U_{AP}M_{P}M_{P}U_{PC}\\
M_{\mathcal{CP}^{-1}} &= U^{'}_{CP}U^{'}_{PA}=U_{CP}M_{P}M_{P}U_{PA}
\end{split}
\end{equation}
where $M_{P}$ is the constraint matrix on object type P (paper).

\subsection{Co-ranking for objects and relations in HIN}

Until now, we have created methods to rank same or different types of objects under a certain constrained meta path. However, there are many constrained meta paths in heterogeneous networks. It is an important issue to automatically determine the importance of paths \cite{SHYYW11,SNHYYY12}, since it is usually hard for us to identify which relation is more important in real applications. To solve this problem (i.e., \emph{Q. 3}), we propose the HRank-CO to co-rank the importance of objects and relations. The basic idea is based on an intuition that important objects are connected to many other objects through a number of important relations and important relations connect many important objects. So we organize the multiple relation networks with a tensor and a random walk process is designed on this tensor. The method not only can comprehensively evaluate the importance of objects by considering all constrained meta paths, but also can rank the contribution of different constrained meta paths.

\begin{figure}[htbp]
\centering
\small
\subfigure[Multiple relations]{\label{fig:fft:a}
\begin{minipage}[t]{0.18\textwidth}
  \includegraphics[width=3.2cm]{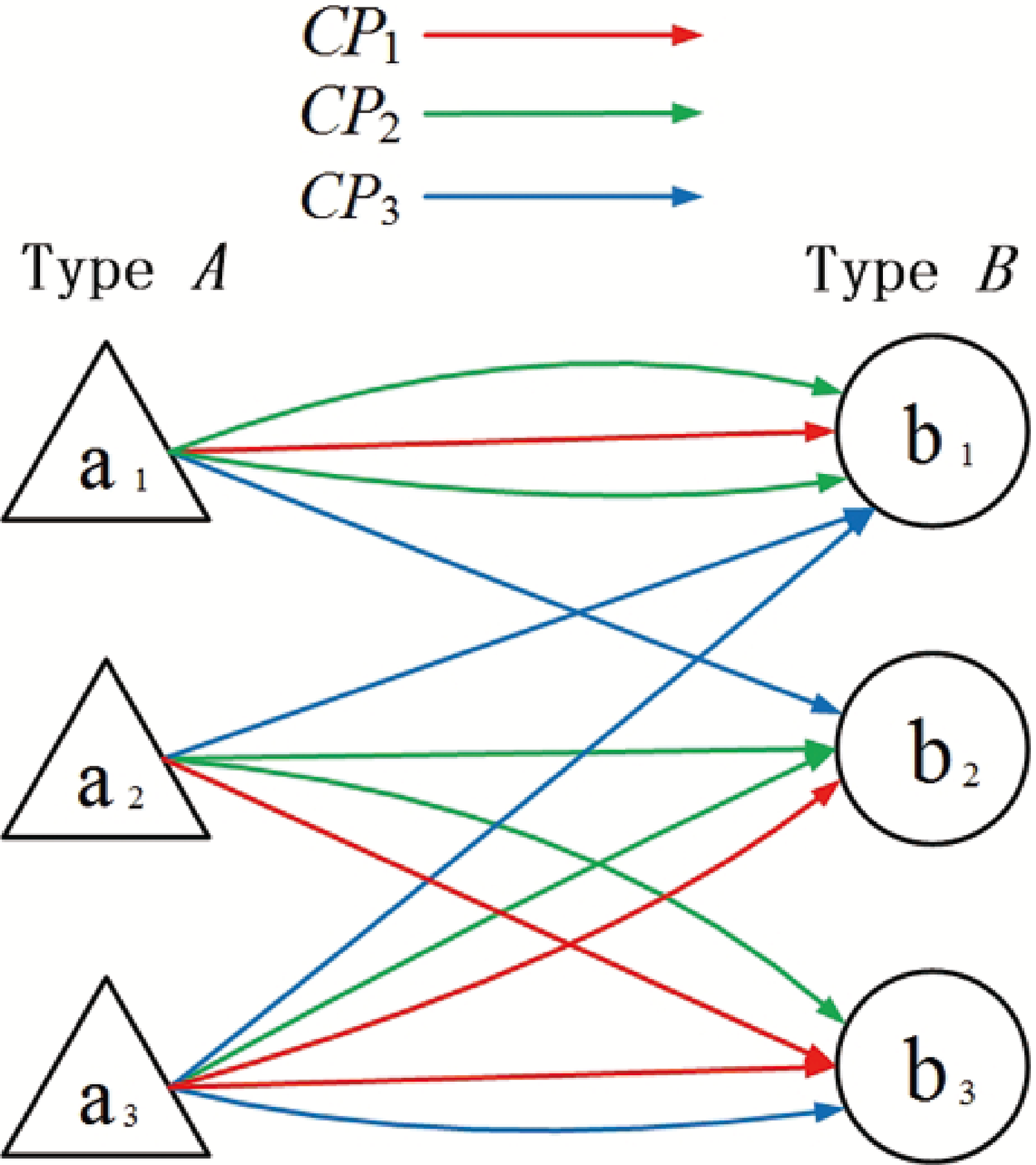}
\end{minipage}%
}%
\subfigure[Tenser representation]{
\begin{minipage}[t]{0.3\textwidth}
  \includegraphics[width=4.7cm]{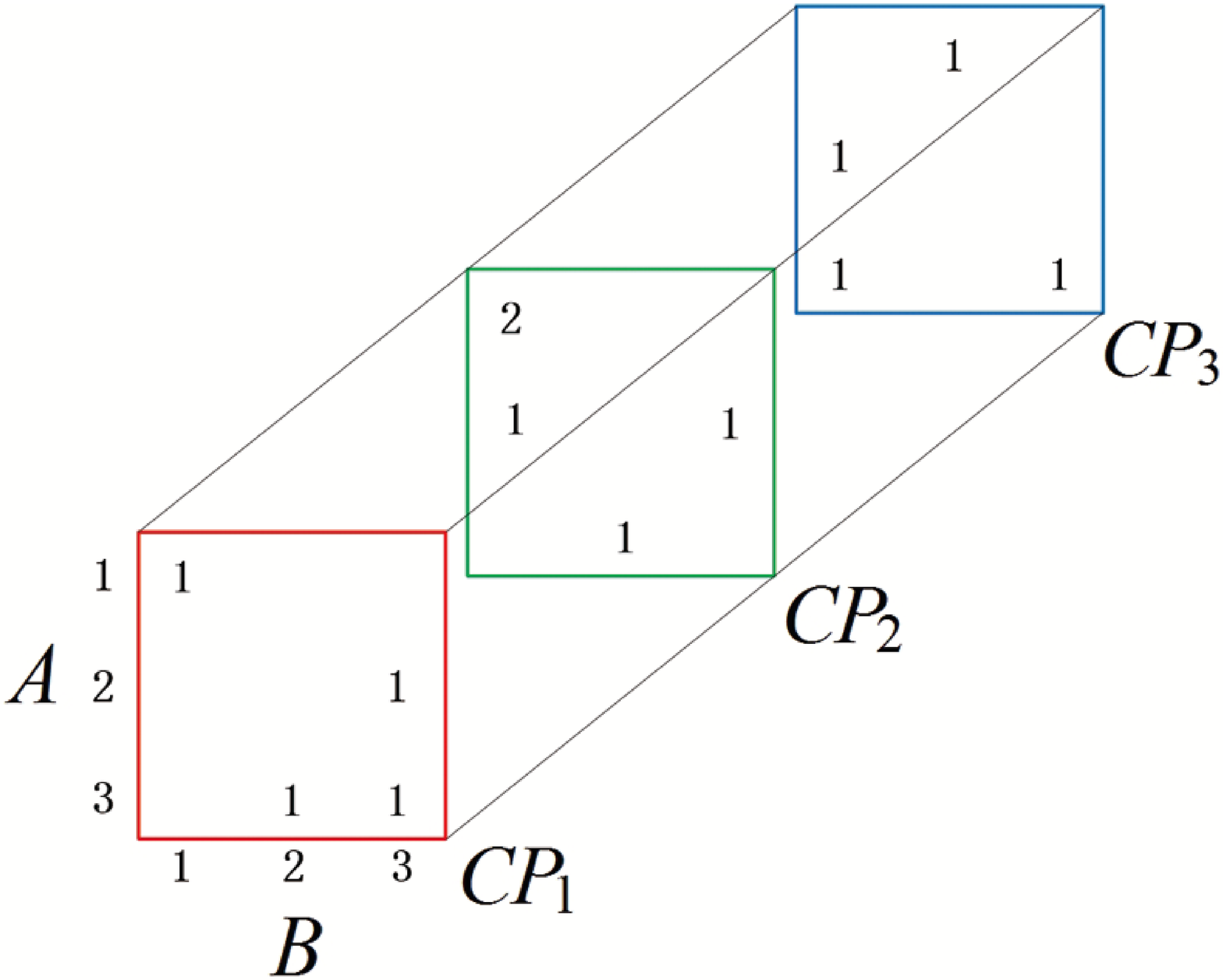}
\end{minipage}
}
\caption{\small An example of multi-relations of objects generated by multiple paths: (a) the graph representation; (b) the corresponding tensor representation.}\label{fig:fft}
\end{figure}

In Figure 3(a), we show an example of multiple relations among objects, generated by multiple meta paths. There are three objects of type $A$, three objects of type $B$ and three types of relations among them. These relations are generated by three constrained meta paths with type $A$ as the source type and type $B$ as the target type. To describe the multiple relations among objects, we use the representation of tensor which is a multidimensional array. We call $X = (x_{i,j,k})$ a 3rd order tensor, where $x_{i,j,k} \in R$, for $i = 1, \cdots, m$, $j = 1, \cdots, l$, $k = 1, \cdots, n$. $x_{i,j,k}$ represents the times that object $i$ is related to object $k$ through the $j$th constrained meta path. For example, Figure 3(b) is a three-way array, where each two dimensional slice represents an adjacency matrix for a single relation. So the data can be represented as a tensor of size $3\times 3\times 3$. In the multi-relational network, we define the transition probability tensor to present the transition probability among objects and relations.

\begin{defn}
\textbf{Transition probability tensor}. In a multi-relational network, $X$ is the tensor representing the network. $F$ is the normalized tensor of $X$ along the column vector. $R$ is the normalized tensor of $X$ along the tube vector. $T$ is the normalized tensor of $X$ along the row vector. $F$, $R$, and $T$ are called the transition probability tensor which can be denoted as follows:
\begin{equation}\small\small
\begin{split}
f_{i,j,k} = \cfrac{x_{i,j,k}}{\sum_{i=1}^{m}x_{i,j,k}}\qquad i = 1,2,\ldots,m
\\
r_{i,j,k} = \cfrac{x_{i,j,k}}{\sum_{j=1}^{l}x_{i,j,k}}\qquad j = 1,2,\ldots,l
\\
t_{i,j,k} = \cfrac{x_{i,j,k}}{\sum_{k=1}^{n}x_{i,j,k}}\qquad k = 1,2,\ldots,n
\end{split}
\end{equation}
\end{defn}
$f_{i,j,k}$ can be interpreted as the probability of object $i$ (of type $A$) being the visiting object when relation $j$ is used and the current object being visited is object $k$ (of type $B$), $r_{i,j,k}$ represents the probability of using relation $j$ given that object $k$ is visited from object $i$, and $t_{i,j,k}$ can be interpreted as the probability of object $k$ being visited, given that object $i$ is currently the visiting object and relation $j$ is used. The meaning of these three tensors can be defined formally as follows:

\begin{equation}\small
\begin{split}
f_{i,j,k} = Prob(X_t = i|Y_t = j, Z_t = k)
\\
r_{i,j,k} = Prob(Y_t = j|X_t = i, Z_t = k)
\\
t_{i,j,k} = Prob(Z_t = k|X_t = i, Y_t = j)
\end{split}
\end{equation}
in which $X_t$, $Z_t$ and $Y_t$ are three random variables representing visiting at certain object of type $A$ or type $B$ and using certain relation respectively at the time $t$.

Now, we define the stationary distributions of objects and relations as follows\\\\
\begin{equation}\small
\begin{split}
x &= (x_1,x_2,\cdots,x_m)^T\\
y &= (y_1,y_2,\cdots,y_l)^T\\
z &= (z_1,z_2,\cdots,z_n)^T
\end{split}
\end{equation}
in which
\begin{equation}\small
\begin{split}
x_i = \lim_{t\rightarrow\infty} Prob(X_t = i)\\
y_j = \lim_{t\to\infty} Prob(Y_t = j)\\
z_k = \lim_{t\rightarrow\infty} Prob(Z_t = k)
\end{split}
\end{equation}

From the above equations, we can get:
\begin{equation}\small
\begin{split}
Prob(X_t = i) = \sum^{l}_{j = 1}\sum^{n}_{k = 1} f_{i,j,k} \times Prob(Y_t = j, Z_t = k)
\\
Prob(Y_t = j) = \sum^{m}_{i = 1}\sum^{n}_{k = 1} r_{i,j,k} \times Prob(X_t = i, Z_t = k)
\\
Prob(Z_t = k) = \sum^{m}_{i = 1}\sum^{l}_{j = 1} t_{i,j,k} \times Prob(X_t = i, Y_t = j)
\end{split}
\end{equation}\\
where $Prob(Y_t = j, Z_t = k)$ is the joint probability distribution of $Y_t$ and $Z_t$, $Prob(X_t = i, Z_t = k)$ is the joint probability distribution of $X_t$ and $Z_t$, and $Prob(X_t = i, Y_t = j)$ is the joint probability distribution of $X_t$ and $Y_t$.

To obtain $x_i$, $y_j$ and $z_k$, we assume that $X_t$, $Y_t$ and $Z_t$ are all independent from each other which can be denoted as below:

\begin{equation}\small
\begin{split}
Prob(X_t = i,Y_t = j) = Prob(X_t = i)Prob(Y_t = j)
\\
Prob(X_t = i,Z_t = k) = Prob(X_t = i)Prob(Z_t = k)
\\
Prob(Y_t = j,Z_t = k) = Prob(Y_t = j)Prob(Z_t = k)
\end{split}
\end{equation}\\

Consequently, through combining the equations with the assumptions above, we get

\begin{equation}\small
\begin{split}
x_i = \sum^{l}_{j = 1}\sum^{n}_{k = 1} f_{i,j,k} y_jz_k,\qquad i = 1,2,\ldots,m,
\\
y_j = \sum^{m}_{i = 1}\sum^{n}_{k = 1} r_{i,j,k} x_iz_k,\qquad j = 1,2,\ldots,l,
\\
z_k = \sum^{m}_{i = 1}\sum^{l}_{j = 1} t_{i,j,k} x_iy_j,\qquad k = 1,2,\ldots,n.
\end{split}
\end{equation}

The equations above can be written in a tensor format:

\begin{equation}\small
x = Fyz,\qquad y = Rxz,\qquad z = Txy
\end{equation}

with\\

$\qquad\qquad\sum_{i = 1}^{m}x_i = 1$, $\sum_{j = 1}^{l}y_j = 1$, and $\sum_{k = 1}^{n}z_k = 1$.

According to the analysis above, we can design the following algorithm to co-rank the importance of objects and relations.

\begin{algorithm}
\caption {HRank-CO Algorithm}
\begin{algorithmic}
\algsetup{linenosize=\small} \small

\REQUIRE Three tensors $F$, $T$ and $R$, three initial probability distributions $x_0$, $y_0$ and $z_0$ and the tolerance $\epsilon$.
\ENSURE Three stationary probability distributions $x$, $y$ and $z$.
\STATE \textbf{Procedure:}
\STATE Set $t = 1$;
\REPEAT
\STATE Compute $x_t = Fy_{t-1}z_{t-1}$;
\STATE Compute $y_t = Rx_{t}z_{t-1}$;
\STATE Compute $z_t = Tx_{t}y_{t}$;
\UNTIL{$||x_{t} - x_{t-1}|| + ||y_{t} - y_{t-1}|| + ||z_{t} - z_{t-1}|| < \epsilon$}
\end{algorithmic}
\end{algorithm}

\subsection{Discussion}
First we analyze the connection of three versions of HRank. We have stated that HRank-SY is a special version of HRank-AS when the asymmetric path degrades to a symmetric path. We can also find that HRank-AS is the special version of HRank-CO. When there is only one relation in HRank-CO, generated by path $\mathcal{P}$, $T$ and $F$ are the transition probability matrices between type $A$ and $B$ along path $\mathcal{P}$ and $\mathcal{P}^{-1}$ (i.e., $M_{\mathcal{P}}$ and $M_{\mathcal{P}^{-1}}$), respectively. Moreover, $R$ and $y$ become 1. In this condition, Eq. 12 turns into Eq. 3 without considering the restarting probability.

Then we estimate the space and time complexity of HRank. For simplicity, we assume that there are $r$ relations, $n$ objects for each type, and $t$ iterations for convergence. For HRank-CO, the space complexity is $O(rn^2)$ to store the transition probability tensor. The time complexity of HRank-CO comes from two parts: iteratively compute rank values (see Algorithm 1) and construct the transition probability tensor (see Def. 8). The time complexity of rank computation is $O(trn^2)$. For the $l$ length path, the complexity of constructing the transition probability tensor is $O(rln^3)$. So the whole time complexity is $O(trn^2+rln^3)$. For HRank-AS and HRank-SY, the number of relations (i.e., $r$) is 1, so their space complexity are $O(n^2)$ and time complexity are $O(tn^2+ln^3)$. For real applications, the relation matrices are very sparse, so the real time complexity is much smaller than the theoretical analysis.

\section{Fast computation Strategies}
According to the time complexity analysis above, we can find that the main time-consuming component of HRank lies in constructing the reachable probability matrix with the complexity $O(rln^3)$. We design three fast computation strategies to fasten the matrix multiplication process to construct the reachable probability matrix.

\subsection{Dynamic Programming Strategy}
Since the matrix multiplication obeys the associative property (i.e.,
$(M_1\times M_2) \times M_3$ is equal to $M_1\times (M_2 \times
M_3)$), we can design the \textbf{Dyn}amic \textbf{P}rogramming
strategy (DynP) to change the sequence of matrix multiplication with the
associative property. The basic idea of DynP is to assign
small-dimensioned matrix with the high computation priority. For a meta
path $\mathcal{P}=R_1\circ R_2\circ \cdots \circ R_{l}$, we can calculate the
expected minimal computation complexity by the following equation and record the computation
sequence in $i$.

\begin{equation}
\tiny
\begin{split}
C(R_1\circ \cdots \circ R_{l}) =
&\left\{ \begin{array}{lll}
0 & \textrm{$l=1$}\\
|R_1.S|\times|R_1.T|\times|R_2.T| & \textrm{$l=2$}\\
\arg\underset{i}{\min}\{C(R_1\cdots R_i)+C(R_{i+1}\cdots R_{l})+\\|R_1.S|\times|R_i.T|\times|R_l.T|\} &\textrm{$l>2$}\\
\end{array} \right.
\end{split}
\end{equation}
Through dynamic programming method, the above equation can be easily solved with the $O(l^2)$ complexity. The running time can be omitted, since $l$ is much smaller than the matrix dimension. The DynP does not change relevance results, so it is an information-lossless strategy.

\subsection{Truncation Strategy}
The basic hypothesis of \textbf{Trun}cation strategy (Trun) is that removing the
probability on those less important nodes would not significantly
degrade the performance. It has been proved by many researches
\cite{LC10a,SC07}. Through keeping matrix sparse, the truncation strategy could greatly reduce the amount of space and time consumption. We can add a
truncation step at each step of the matrix multiplication. In the truncation step,
the probability value is set with 0 for those nodes whose
values are smaller than a threshold $\varepsilon$. Although a
static threshold is usually used in many methods (e.g., ref. \cite{LC10a}), it faces the following disadvantage: it may truncate nothing for matrix whose elements all have high probability and it may truncate most nodes for matrix whose elements all have low probability. Since we usually pay close attention to the top $k$
objects in query task, the threshold $\varepsilon$ can be set as the
top $k$ relevance value for each search object. The $k$ is dynamically
adjusted as follows.
\begin{displaymath}\small
k = \left\{ \begin{array}{ll}
L & \textrm{if $L\leq W$}\\
\lfloor {(L-W)}^{\beta}\rfloor+W (\beta\in[0,1]) & \textrm{otherwise}\\
\end{array} \right.
\end{displaymath}
where $L$ is the vector length and $W$ is the number of top objects,
decided by users. The $W$ and $\beta$ determine the truncation
level. The larger $W$ or $\beta$ will cause the larger $k$, which
means a denser matrix. It is expensive to determine the top $k$
relevance value for each object, so we can estimate the value by the top $kM$
value for the whole matrix ($M$ is the number of objects). However, it is
also time-consuming to calculate the top $kM$ value for the whole
matrix. It can be approximated with the sample data from the raw
matrix. The sample ratio is $\gamma$. The larger $\gamma$ leads to
more accurate approximation with longer running time. In summary,
the truncation strategy is an information-loss strategy. It can keep
matrix sparse with small sacrifice on accuracy. In addition, it
needs additional time to estimate the threshold $\varepsilon$.

\subsection{Monte Carlo Strategy}
The basic idea of \textbf{Mon}te \textbf{C}arlo method (MonC) is that, for each node $u$, $K$ independent random walkers are simulated
starting from $u$. The distribution of \emph{u} is
approximated by the normalized counts of number of times
the random walkers visit a node. So the reachable probability $PM_{\mathcal{P}}(a,b)$ can be approximated by the normalized count of the number of times that the walkers visit the node
$b$ from $a$ along the path $\mathcal{P}$.
\begin{displaymath}\small
PM_{\mathcal{P}}(a,b)=\frac{\# times\ the \ walkers\  visit\  b\  along\  \mathcal{P}}{\# walkers \ from \ a}
\end{displaymath}

The number of walkers from $a$ (i.e., $K$) controls the accuracy and amount of computation. The larger $K$ will achieve more accurate estimation with more time cost. An advantage of the MonC strategy is that its running time is not affected by the dimension and sparsity of matrix. However, the high-dimension matrix needs larger $K$ for high accuracy. As a sampling method, the MonC is also an information-loss strategy.

%\vspace{-10pt}
\begin{figure}[htbp]
  \centering
\subfigure[DBLP]{\label{fig:fft:a}
\begin{minipage}[t]{0.12\textwidth}
  \includegraphics[width=2cm]{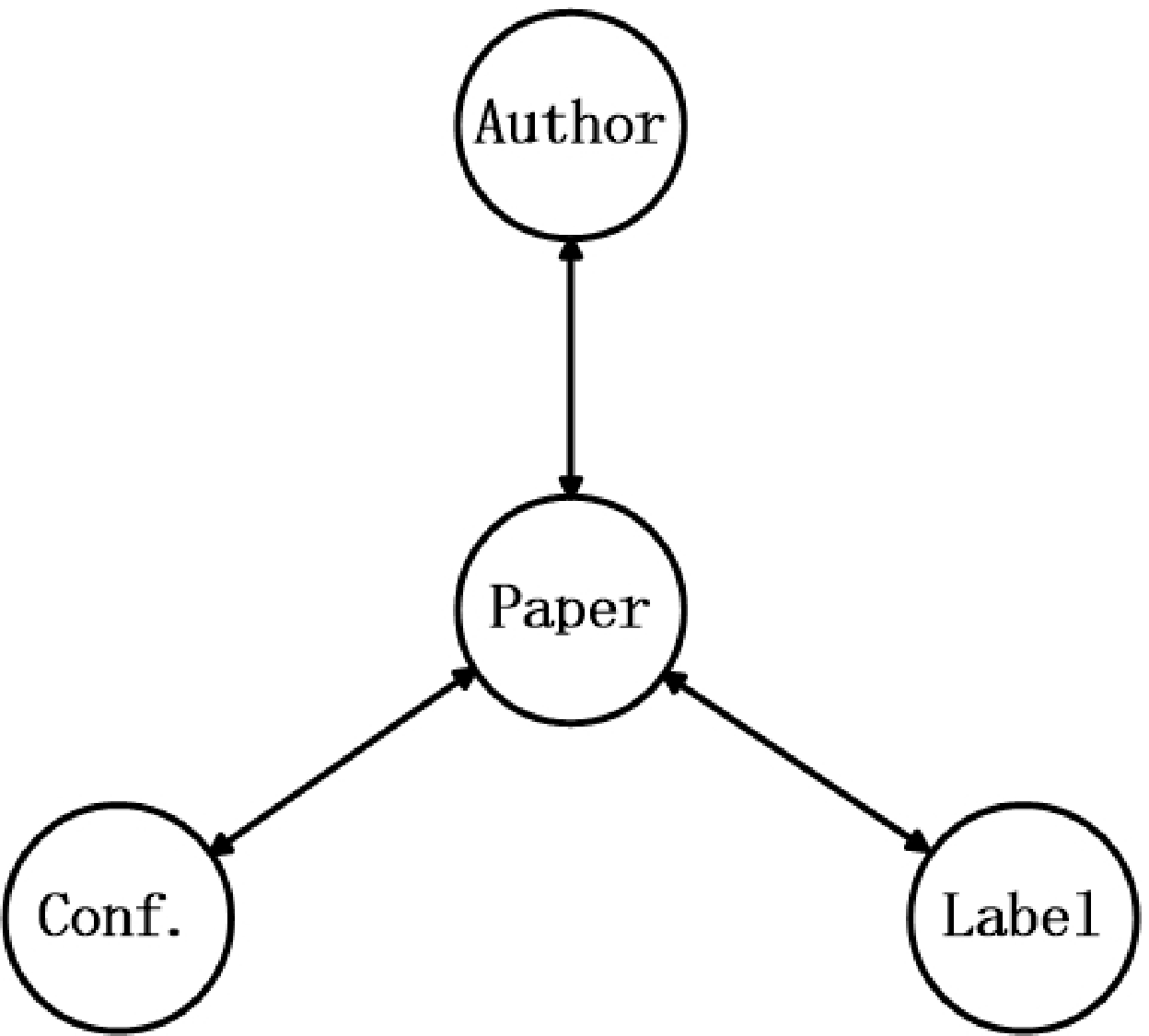}
\end{minipage}%
}%
\subfigure[ACM]{
\begin{minipage}[t]{0.18\textwidth}
  \includegraphics[width=3cm]{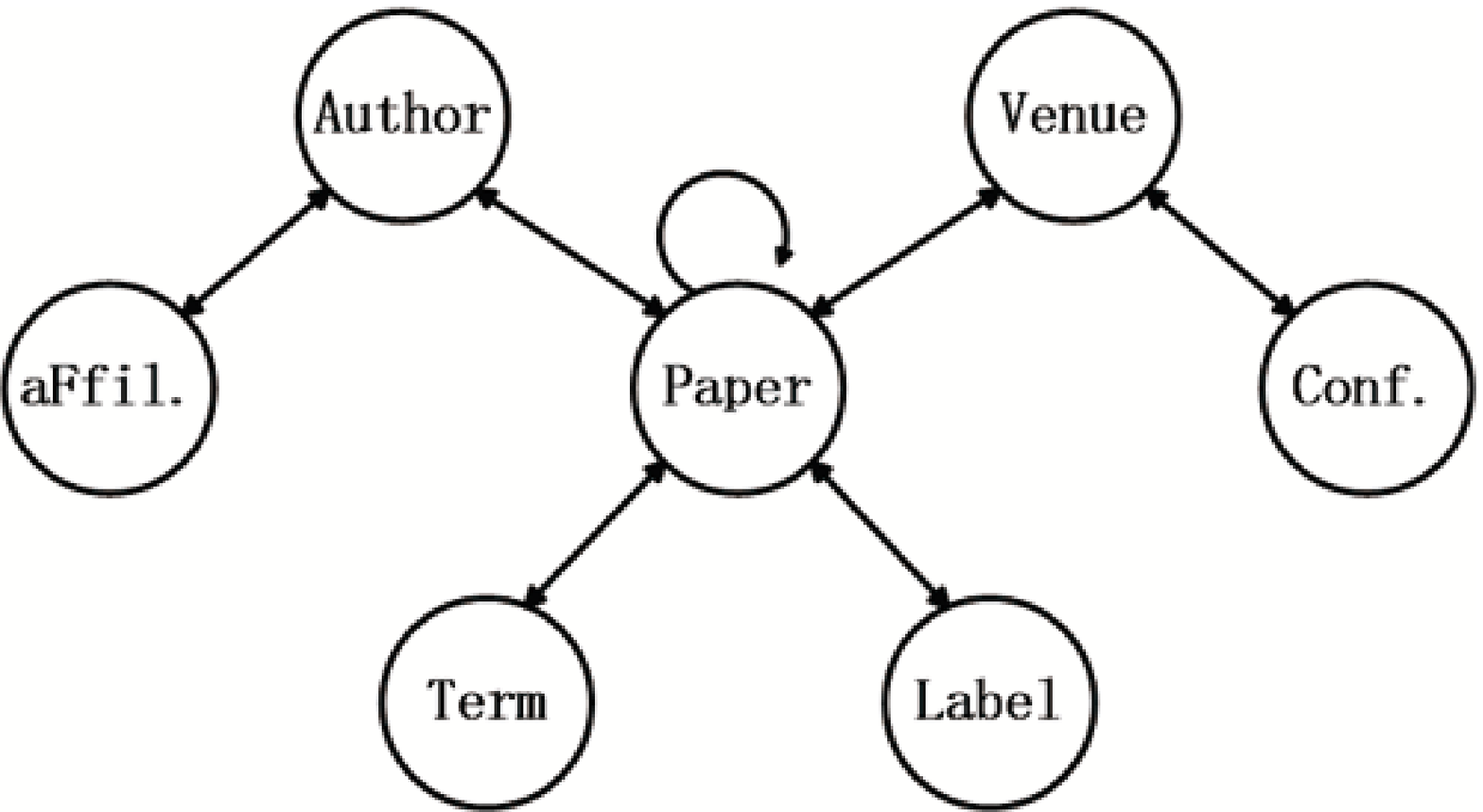}
\end{minipage}
}%
\subfigure[Movie]{
\begin{minipage}[t]{0.12\textwidth}
  \includegraphics[width=2cm]{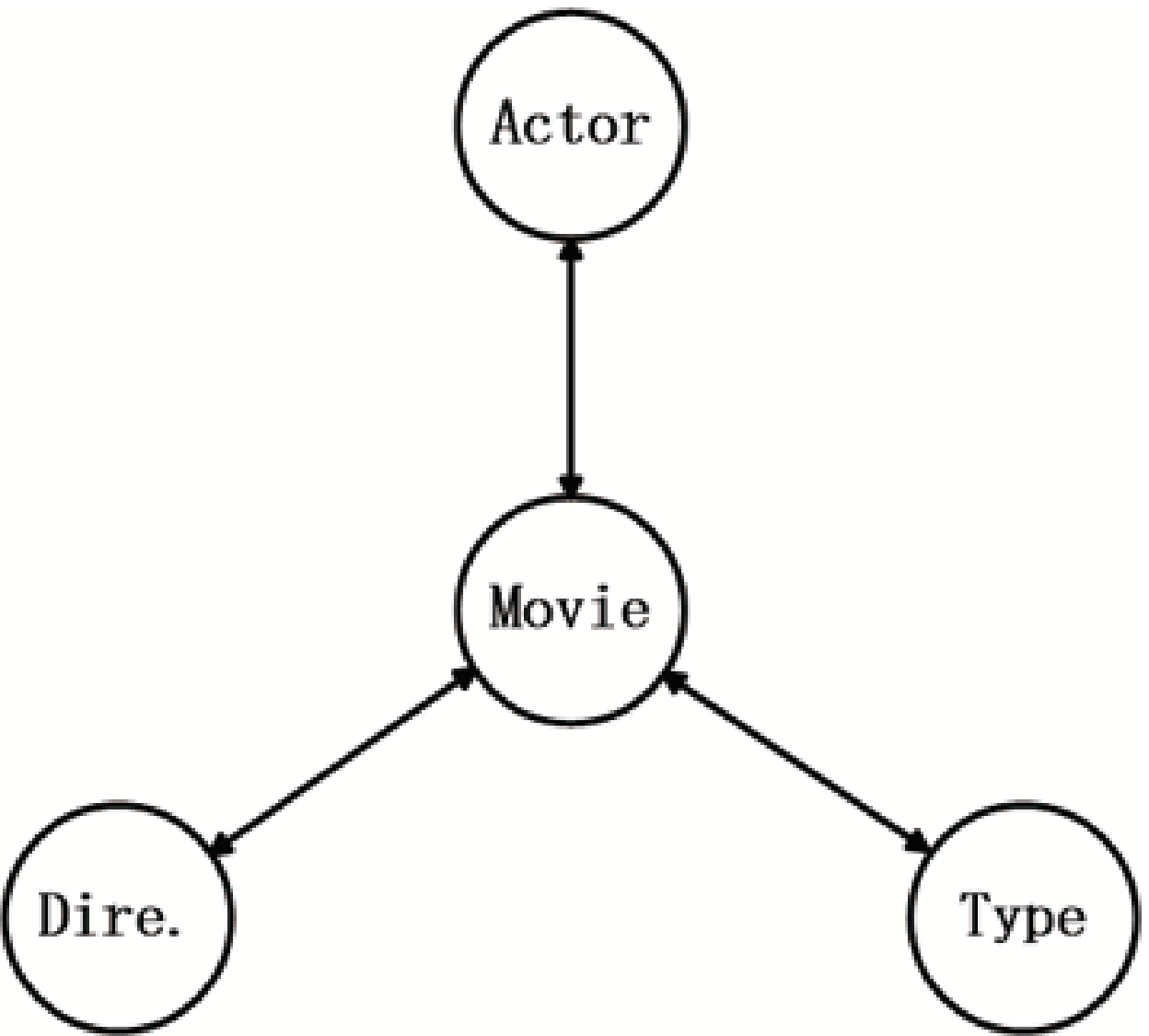}
\end{minipage}
}%\vspace{-5pt}
\caption{\small The network schema of three heterogeneous datasets. (a) DBLP bibliographic dataset. (b) ACM bibliographic dataset. (c) IMDB movie dataset.}\label{fig:schema}
\end{figure}
%\vspace{-10pt}

\section{Experiments}
In this section, we do experiments to validate the effectiveness of three versions of HRank on three real datasets, respectively.

\subsection{Datasets}
We use three heterogeneous information networks for our experiments, including DBLP dataset, ACM dataset, and IMDB dataset. They are summarized as follows:

\textbf{DBLP dataset} \cite{SKYX12,SHYYW11}: The DBLP dataset is a sub-network collected from DBLP website \footnote{http://www.informatik.uni-trier.de/$\sim$ley/db/} involving major conferences in two research areas: database (DB) and information retrieval (IR), which naturally form two labels. The dataset contains 9682 authors, 20 conferences and 22185 papers which are all labeled with one of the two research areas. The network schema is shown in Figure 4(a).

\textbf{ACM dataset} \cite{SKYX12}: The ACM dataset was downloaded from ACM digital library \footnote{http://dl.acm.org/} in June 2010. The ACM dataset comes from 14 representative computer science conferences: KDD, SIGMOD, WWW, SIGIR, CIKM, SODA, STOC, SOSP, SPAA, SIGCOMM, MobiCOMM, ICML, COLT and VLDB. These conferences include 196 corresponding venue proceedings (e.g., KDD conference includes 12 proceedings, such as KDD'10, KDD'09, etc). The dataset has 12499 papers, 17431 authors, 1903 terms and 1804 author affiliations. The network also includes 73 labels of these papers in ACM category (e.g., H.2 is Database Management). The network schema of ACM dataset is shown in Figure 4(b).

\textbf{IMDB dataset} \cite{SZKYLW12}: We crawled movie information from The Internet Movie Database \footnote{www.imdb.com/} to construct the network. The IMDB movie data collects 1591 movies before 2010. The related objects include movies,
actors, directors and movie types, which are organized as a star schema shown in Figure 4(c). Movie information includes 5324 actors, 1591 movies, 551 directors and 112 movie types (e.g., comedy and romance).

\begin{table*}[htbp]
\caption{Top ten authors of different methods on ACM dataset. The number in the parenthesis of the fifth column means the rank of authors in the whole ranking list returned by PageRank. }
\begin{center}
\resizebox{.7\textwidth}{!}{
\begin{tabular}{|c|c|c|c|c|c|}
\hline
Rank & \multicolumn{1}{c|}{$APA$} & \multicolumn{1}{c|}{$APA|P.L = ``H.3"$} & \multicolumn{1}{c|}{$APA|P.L = ``H.2"$} & \multicolumn{1}{c|}{PageRank}& \multicolumn{1}{c|}{Degree}\\\hline
1& {Jiawei Han}& {W. Bruce Croft}& {Jiawei Han}& {Ming Li(1522)}& {Jiawei Han}   \\
2& {Philip Yu}& {ChengXiang Zhai}& {Christos Faloutsos} & {Wei Wei(2072)}& {Philip Yu} \\
3& {Christos Faloutsos}& {James Allan}& {Philip Yu} & {Jiawei Han(5385)}& {ChengXiang Zhai} \\
4& {Zheng Chen}& {Jamie Callan}& {Jian Pei}& {Tao Li(6090)}& {Zheng Chen}  \\
5& {Wei-Ying Ma}& {Zheng Chen}& {H. Garcia-Molina}& {Hong-Jiang Zhang(6319)}& {Christos Faloutsos}  \\
6& {ChengXiang Zhai}& {Ryen W. White}& {Jeffrey F. Naughton}& {Wei Ding(6354)}& {Ravi Kumar}  \\
7& {W. Bruce Croft}& {Wei-Ying Ma}& {Divesh Srivastava}  & {Jiangong Zhang(7285)}& {W. Bruce Croft}\\
8& {Scott Shenker}& {Jian-Yun Nie}& {Raghu Ramakrishnan} & {Christos Faloutsos(7895)}& {Wei-Ying Ma} \\
9& {H. Garcia-Molina}& {Gerhard Weikum}& {Charu C. Aggarwal} & {Feng Pan(8262)}& {Gerhard Weikum} \\
10& {Ravi Kumar}& {C. Lee Giles}& {Surajit Chaudhuri}& {Hongyan Liu(8440)}& {Divesh Srivastava} \\\hline
\end{tabular}
}
\end{center}
\end{table*}

\subsection{Ranking of Homogeneous Objects}
Since the homogeneous objects are connected by symmetric constrained meta paths, the experiments validate the effectiveness of HRank-SY on symmetric constrained meta paths.

\subsubsection{Experiment Study on Symmetric Constrained Meta Paths}
This experiment ranks same type of objects by designating a symmetric constrained meta path on ACM dataset. Here we rank the importance of authors through the symmetric meta path $APA$, which considers the co-author relations among authors. We also employ two constrained meta paths $APA|P.L = ``H.2"$ and $APA|P.L = ``H.3"$, where the categories of ACM $H.2$ and $H.3$ represent ``database management'' and ``information storage/retrieval'', respectively. That is, two constrained meta paths subtly consider the co-author relations in database/data mining field and information retrieval field, respectively. We employ HRank-SY to rank the importance of authors based on these three paths. As the baseline methods, we rank the importance of authors with PageRank and the degree of authors (called Degree method). We directly run PageRank on the whole ACM network by ignoring the heterogeneity of objects. Since the results of PageRank mix all types of objects, we select the author type from the ranking list as the final results.

The top ten authors of each method are shown in Table 1. We can find that these ranking lists all have some common influential authors except that of PageRank. The results of PageRank include some not very well known authors in DB/IR field, such as Ming Li and Wei Wei, although they may be very influential in other fields. We know that the PageRank values of objects are decided by their degrees to a large extent, so the rank values of affiliation objects are high due to their high degrees. It improves the rank values of author objects connecting multiple high-ranking affiliations. The bad results of PageRank show that the ranking in heterogeneous networks should consider the heterogeneity of objects. Otherwise, it cannot distinguish the effect of different types of links. Moreover, we can also observe that the results of HRank with constrained meta paths have obvious bias on the field it assigns. For example, the path $APA|P.L = ``H.3"$ reveals the important authors in information retrieval field, such as W. Bruce Croft, ChengXiang Zhai, and James Allan. However, the path $APA|P.L = ``H.2"$ returns the influential authors in database and data mining field, such as Jiawei Han and Christos Faloutsos. For the meta path $APA$, it mingles well-known authors in these two fields. The results illustrate that the constrained meta paths are able to capture subtle semantics by deeply disclosing the most influential authors in a certain field.

\subsubsection{Quantitative Comparison Experiments}

Based on the results returned by five methods, we can obtain five candidate ranking lists of authors in ACM dataset. To evaluate the results quantitatively, we use the author ranks from Microsoft Academic Search \footnote{http://academic.research.microsoft.com/} as ground truth. Specifically, we crawled two standard ranking lists of authors in two academic fields: DB and IR. Then we compare the difference between our candidate ranking lists and the standard ranking lists. We use the \emph{Distance} criterion \cite{NZWM05} to measure the quality of the ranking results. The criterion not only measures the number of mismatches between these two lists, but also considers the position of these mismatches. The smaller \emph{Distance} means the smaller difference (i.e., better performance).

%\vspace{-10pt}
\begin{figure}[htbp]
\subfigure[DB field]{
\begin{minipage}[t]{0.25\textwidth}
  \includegraphics[width=4.2cm]{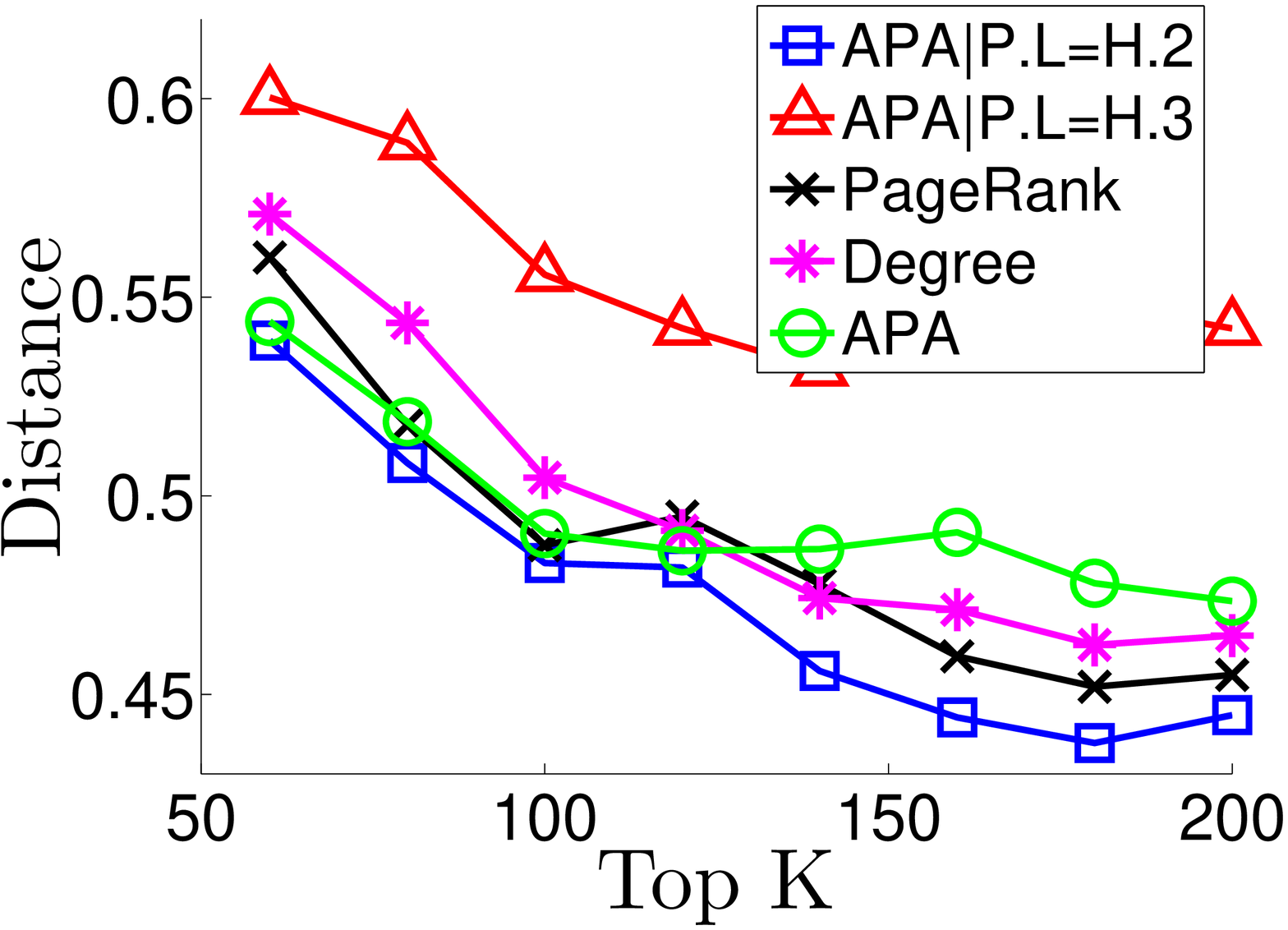}
\end{minipage}
}%
\subfigure[IR field]{
\begin{minipage}[t]{0.25\textwidth}
  \includegraphics[width=4.2cm]{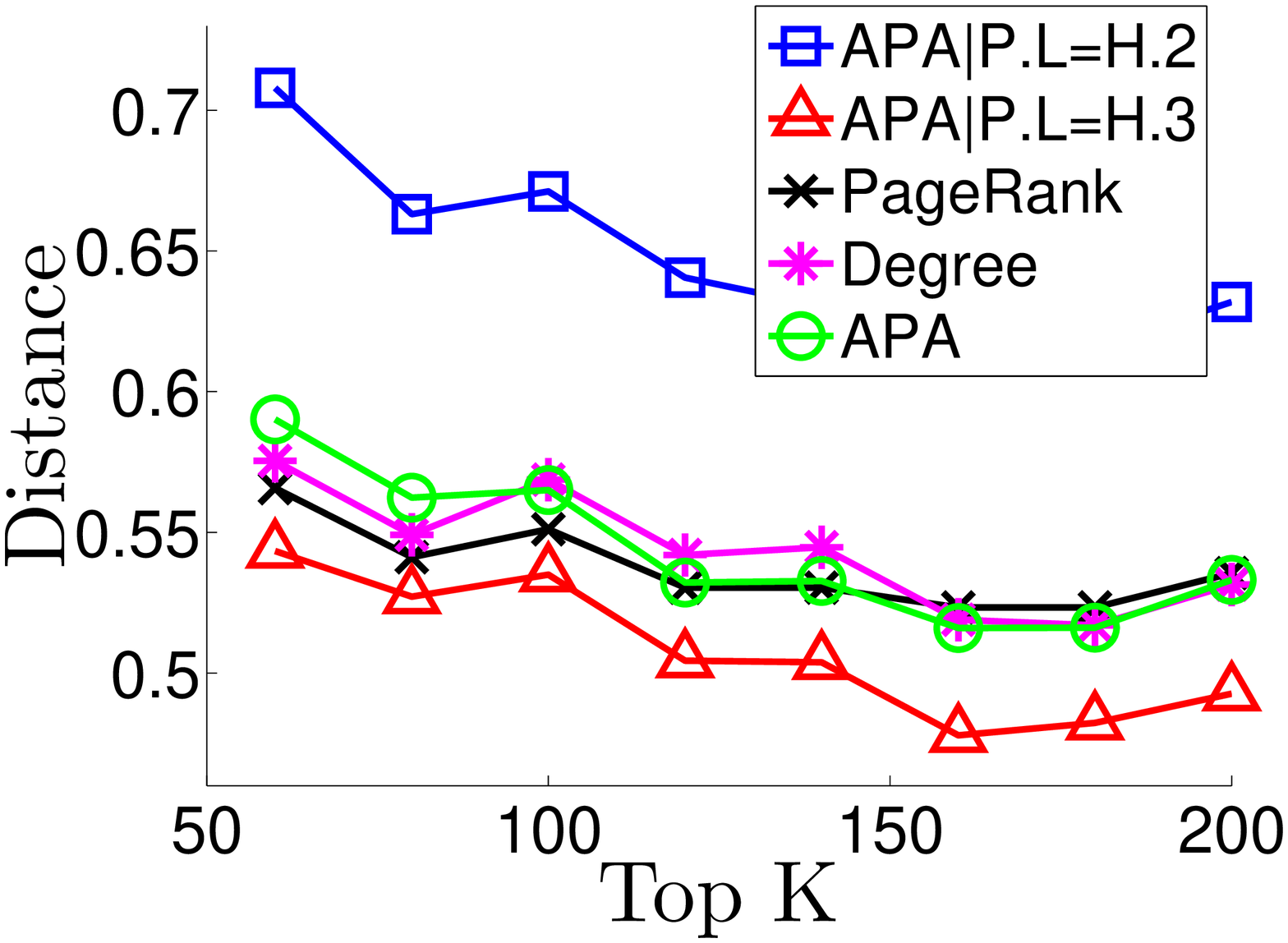}
\end{minipage}
}%
%\subfigure[Whole field]{
%\begin{minipage}[t]{0.15\textwidth}
%  \includegraphics[width=3cm]{./picture/acm_apa_intenew.eps}
%\end{minipage}
%}
%\vspace{-5pt}
\caption{\small The Distances between the ranking lists obtained by different methods and the standard ranking lists on different fields on ACM dataset.}\label{fig:fft}
\end{figure}
%\vspace{-10pt}

In this experiment, we compare the five candidate ranking lists with each of the two standard ranking lists from Microsoft Academic Search and the \emph{Distance} results are shown in Figure 5. We can observe an obvious phenomenon: the results obtained by the constrained meta paths have the smallest \emph{Distance} on its corresponding field, while they have the largest \emph{Distance} on other fields. For example, HRank with the path $APA|P.L = ``H.2"$ has the smallest \emph{Distance} on the DB field in Figure 5(a), while it has the largest \emph{Distance} on IR field in Figure 5(b). The reason lies in that the path $APA|P.L = ``H.2"$ focuses on the authors in the DB field. Meanwhile, these authors deviate from those in the IR field. The results further illustrate that the constrained meta path can disclose the influential authors in a certain field more correctly. Since the meta path (i.e., $APA$) considers the co-author relationship on all fields, it achieves mediocre performances on these two fields. In fact, the HRank with meta path $APA$ only achieves closer performances to PageRank and Degree methods. It implies that the constrained meta path in HRank indeed helps to improve the ranking performances in a specific field.

\begin{table*}[htbp]
\caption{Top ten authors of different methods on DBLP dataset. The number in the parenthesis of the fifth column means the rank of authors in the whole ranking list returned by PageRank.}
\begin{center}
\resizebox{.8\textwidth}{!}{
\begin{tabular}{|c|c|c|c|c|c|}
\hline
Rank & \multicolumn{1}{c|}{$APC$} & \multicolumn{1}{c|}{$APC|P.L=``DB"$} & \multicolumn{1}{c|}{$APC|P.L=``IR"$} & \multicolumn{1}{c|}{PageRank} & \multicolumn{1}{c|}{Degree}\\\hline
1& {Gerhard Weikum}& {Surajit Chaudhuri}& {W. Bruce Croft} & {W. Bruce Croft(23)} & {Philip S. Yu}   \\
2& {Katsumi Tanaka}&  {H. Garcia-Molina}& {Bert R. Boyce} & {Gerhard Weikum(24)} & {Gerhard Weikum}  \\
3& {Philip S. Yu}&  {H. V. Jagadish}& {Carol L. Barry} & {Philip S. Yu(25)} & {Divesh Srivastava}  \\
4& {H. Garcia-Molina}&  {Jeffrey F. Naughton}& {James Allan} & {Jiawei Han(26)} & {Jiawei Han}  \\
5& {W. Bruce Croft}&  {Michael Stonebraker}& {ChengXiang Zhai} & {H. Garcia-Molina(27)} & {H. Garcia-Molina}  \\
6& {Jiawei Han}&  {Divesh Srivastava}& {Mark Sanderson} & {Divesh Srivastava(28)} & {W. Bruce Croft}  \\
7& {Divesh Srivastava}& {Gerhard Weikum}& {Maarten de Rijke} & {Surajit Chaudhuri(29)} & {Surajit Chaudhuri}  \\
8& {Hans-Peter Kriegel}&  {Jiawei Han}& {Katsumi Tanaka} & {H. V. Jagadish(30)} & {H. V. Jagadish}  \\
9& {Divyakant Agrawal}& {Christos Faloutsos}& {Iadh Ounis} & {Jeffrey F. Naughton(31)} & {Jeffrey F. Naughton} \\
10& {Jeffrey Xu Yu}& {Philip S. Yu}& {Joemon M. Jose} & {Rakesh Agrawal(32)} & {Rakesh Agrawal}  \\\hline
\end{tabular}
}

\end{center}
\end{table*}

\subsection{Ranking of Heterogeneous Objects}
Then the experiments validate the effectiveness of HRank-AS on asymmetric constrained meta paths.

\subsubsection{Experiment Study on Asymmetric Constrained Meta Paths}
The experiments are done on the DBLP dataset. We evaluate the importance of authors and conferences simultaneously based on the meta path $APC$, which means authors publish papers on conferences. Two constrained meta paths ($APC|P.L = ``DB"$ and $APC|P.L = ``IR"$) are also included, which means authors publish DB(IR)-field papers on conferences. Similarly, the experiments also include two baseline methods (i.e., PageRank and Degree) in above experiments with the same experimental process.

The top ten authors and conferences returned by these five methods are shown in Tables 2 and 3, respectively. As shown in Table 2, the ranking results of these methods on authors all are reasonable, however, the constrained meta paths can find the most influential authors in a certain field. For example, the top three authors of $APC|P.L = ``DB"$ are Surajit Chaudhuri, Hector Garcia-Molina and H. V. Jagadish, and all of them are very influential researchers in the database field. The top three authors of $APC|P.L = ``IR"$ are W. Bruce Croft, Bert R. Boyce and Carol L. Barry, and they all have the high academic reputation in information retrieval field. Similarly, as we can see in Table 3, HRank with constrained meta paths (i.e., $APC|P.L = ``DB"$ and $APC|P.L = ``IR"$) can clearly find the important conferences in DB and IR fields, while other methods mingle these conferences. For example, the most important conferences in the DB field are ICDE, VLDB and SIGMOD, while the most important conferences in the IR field are SIGIR, WWW, CIKM. Observing Tables 2 and 3, we can also find the mutual effect of authors and conferences. That is, an influential author published many papers in the important conferences, and vice versa. For example, W. Bruce Croft published many papers in SIGIR and CIKM, while Surajit Chaudhuri has many papers in SIGMOD, ICDE and VLDB.

\begin{table}[htbp]
\caption{\small Top ten conferences of different methods on DBLP dataset. The number in the parenthesis of the fifth column means the rank of conferences in the whole ranking list returned by PageRank.}
\begin{center}
\resizebox{.5\textwidth}{!}{
\begin{tabular}{|c|c|c|c|c|c|}
\hline
Rank & \multicolumn{1}{c|}{$APC$} & \multicolumn{1}{c|}{$APC|P.L=``DB"$} & \multicolumn{1}{c|}{$APC|P.L=``IR"$} & \multicolumn{1}{c|}{PageRank} & \multicolumn{1}{c|}{Degree}\\\hline
1& {CIKM}& {ICDE}& {SIGIR} & {ICDE(3)}& {ICDE}  \\
2& {ICDE}&  {VLDB}& {WWW} & {SIGIR(4)}& {SIGIR} \\
3& {WWW}&  {SIGMOD}& {CIKM}  & {VLDB(5)}& {VLDB}\\
4& {VLDB}&  {PODS}& {JASIST} & {CIKM(6)}& {SIGMOD} \\
5& {SIGMOD}&  {DASFAA}& {WISE} & {SIGMOD(7)}& {CIKM} \\
6& {SIGIR}&  {EDBT}& {ECIR}& {JASIST(8)}& {JASIST}  \\
7& {DASFAA}& {ICDT}& {APWeb}& {WWW(9)}& {WWW}  \\
8& {JASIST}&  {MDM}& {WSDM} & {DASFAA(10)}& {PODS} \\
9& {WISE}& {WebDB}& {JCIS} & {PODS(11)}& {DASFAA} \\
10& {EDBT}& {SSTD}& {IJKM} & {JCIS(12)}& {EDBT}\\\hline
\end{tabular}
}

\end{center}
\end{table}

\subsubsection{Quantitative Comparison Experiments}
To verify the effectiveness of these methods, we use the above \emph{Distance} criterion to calculate the difference between their results and standard ranking lists crawled from Microsoft Academic Search. Figure 6 shows the differences of author ranking lists. We can observe the same phenomenon with above quantitative experiments. That is, HRank with constrained meta paths achieve the best performances on their corresponding field, while they have the worst performances on other fields. In addition, compared to that of PageRank and Degree, the mediocre performances of HRank with meta path $APC$ further demonstrate the importance of constrained meta path to capture the subtle semantics contained in heterogeneous networks.

\begin{figure}[htbp]
\subfigure[DB field]{
\begin{minipage}[t]{0.21\textwidth}
  \includegraphics[width=4cm]{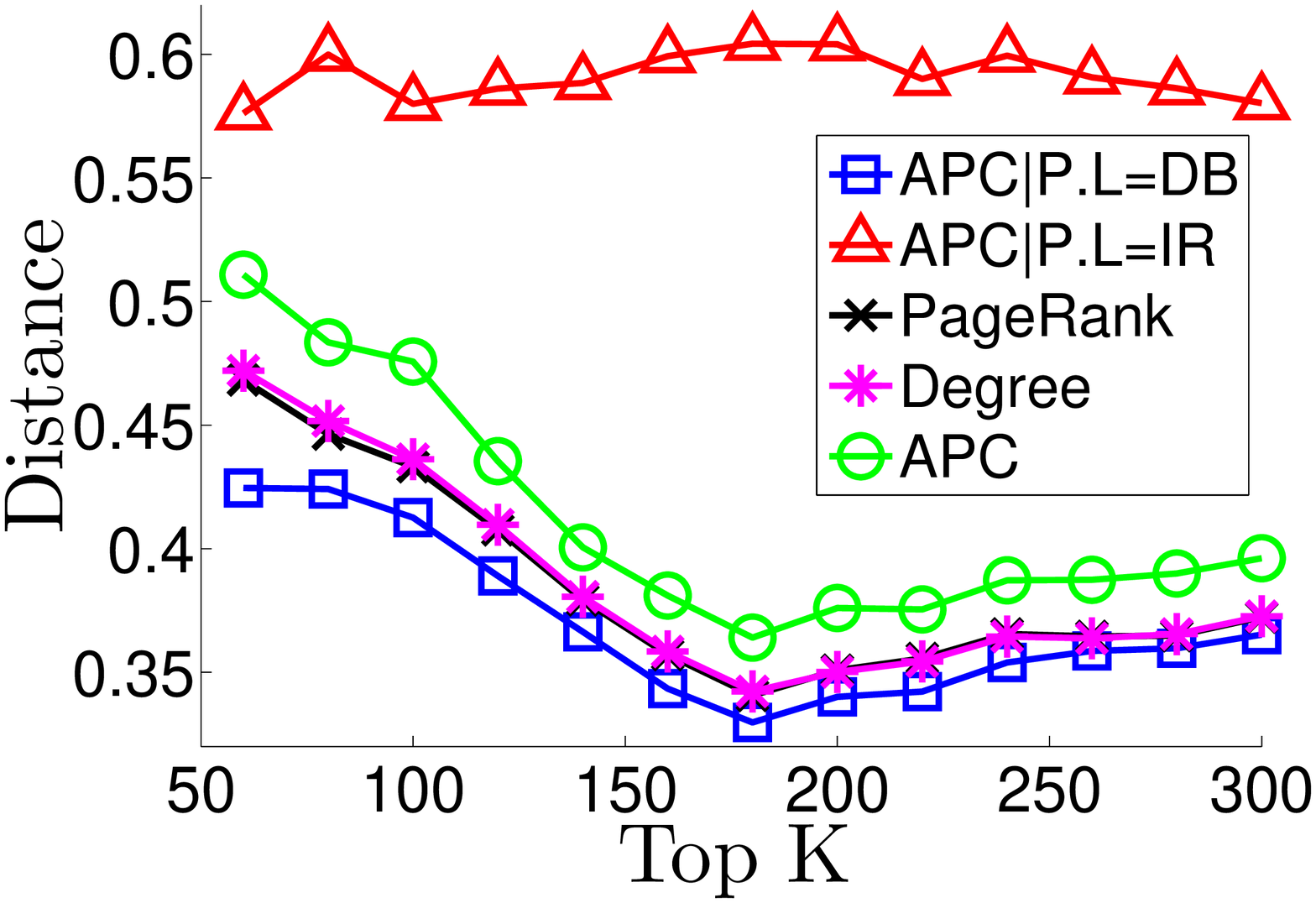}
\end{minipage}
}%
\subfigure[IR field]{
\begin{minipage}[t]{0.3\textwidth}
  \includegraphics[width=4cm]{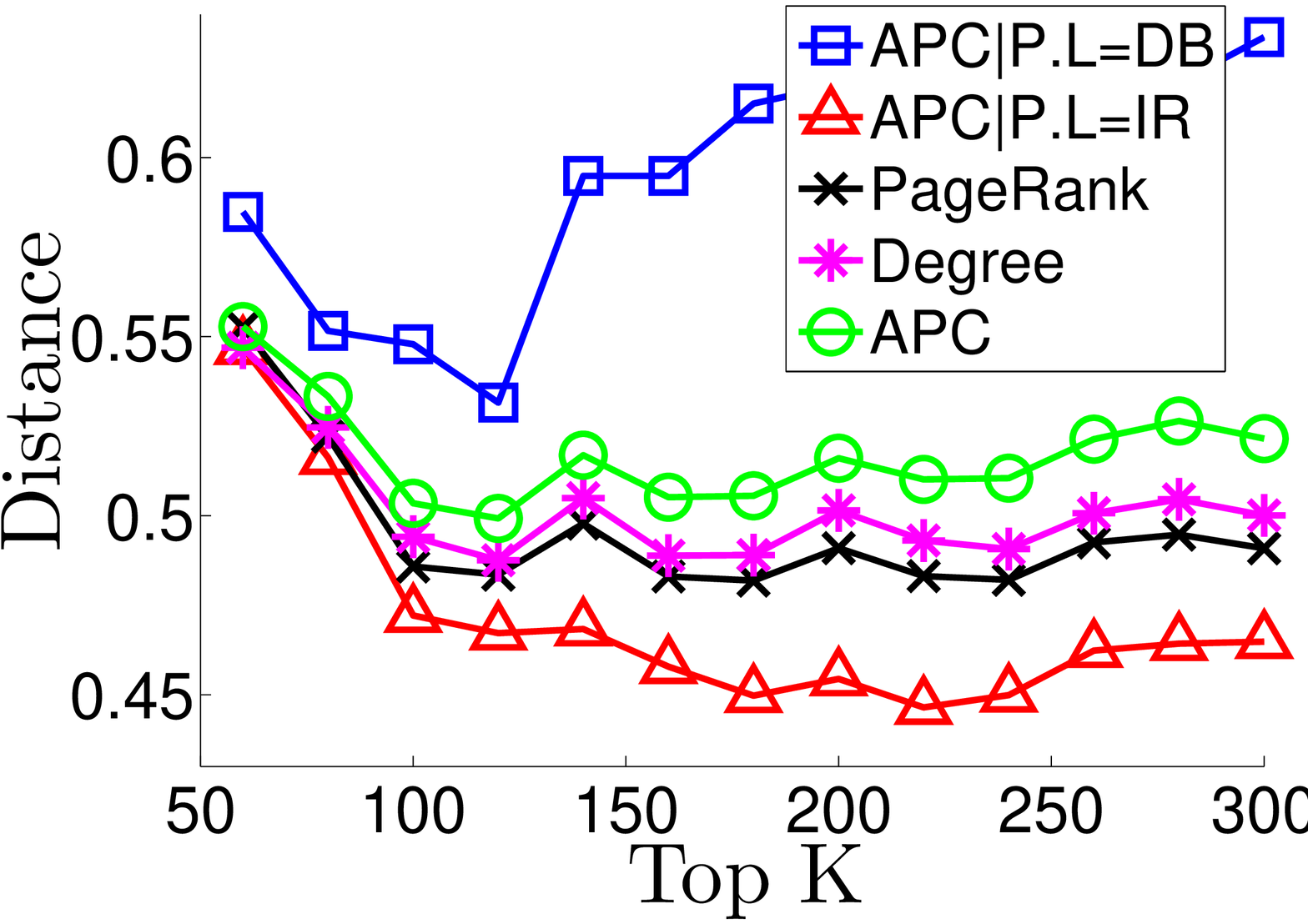}
\end{minipage}
}
\caption{\small The Distances between the candidate author ranking lists and the standard ranking lists on different fields on DBLP dataset.}\label{fig:fft}
\end{figure}

\subsubsection{Experiments on Meta Path with Multiple Constraints}
Furthermore, we validate the effectiveness of meta path with multiple constraints. In the above experiments, we employ the constraint on the label of papers in HRank with the meta path $APC$. Here we add one more constraint on conference. Specifically, by contrast to the constrained meta path $APC|P.L=``DB"$, we employ the paths $APC|P.L=``DB"\&\&C=``VLDB"$, $APC|P.L=``DB"\&\&C=``SIGIR"$, and $APC|P.L=``DB"\&\&C=``CIKM"$, which mean authors publish DB-field papers on specified conferences (e.g., VLDB, SIGIR, and CIKM). Similarly, we add the same conference constraints on the path $APC|P.L=``IR"$. Same with the above experiments, we calculate the rank accuracy of HRank with these constrained meta paths and the results are shown in Figure 7.

We know that HRank with the path $APC|P.L=``DB"$ ($APC|P.L=``IR"$) can reveal the influence of authors in the DB (IR) field. As ground truth, this ranking is based on the aggregate of many conferences related to the DB field. The added conference constraint in HRank further reveals the influence of authors in the specific conference of the field. So we can use the closeness to the ground truth to reveal the importance of a conference to that field. That is, if the ranking from a specific conference is quite closer to the ground truth rank, that can imply the conference is a dominating conference in that field. From Figure 7(a), we can find that the VLDB conference constraint (the blue curve) achieves the closest performances to the ground truth ranking, while the performances of the SIGIR conference constraint (the black curve) deviate most. So we can infer that the VLDB is more important than SIGIR in the DB field and the CIKM has the middle importance. Similarly, from Figure 7(b), we can infer that the SIGIR is more important than VLDB in the IR field. These findings comply with our common sense. As we know, although the VLDB and SIGIR both are the top conferences in computer science, they are very important only in their research fields. For example, the VLDB is important in the DB field, while it is not so important in the IR field. The middle importance of the CIKM conference stems from the fact that it is a comprehensive conference including papers from DB and IR fields. In addition, we can find that the SIGIR curve almost overlaps with the ground truth over the IR field, while the VLDB curve still has a gap with the ground truth over the DB field. We think the reason is that SIGIR is the main conference in the IR field, while in the DB field, there are also other important conferences, such as SIGMOD and ICDE. In all, the experiments show that HRank with constrained meta path can not only effectively find the influential authors in each research field on a specified conference but also indirectly reveal the importance of conferences in fields. It also implies that HRank can achieve accurate and subtle ranking results by flexibly setting the combination of constraints.

\begin{figure}[htbp]
\subfigure[DB field]{
\begin{minipage}[t]{0.21\textwidth}
  \includegraphics[width=4cm]{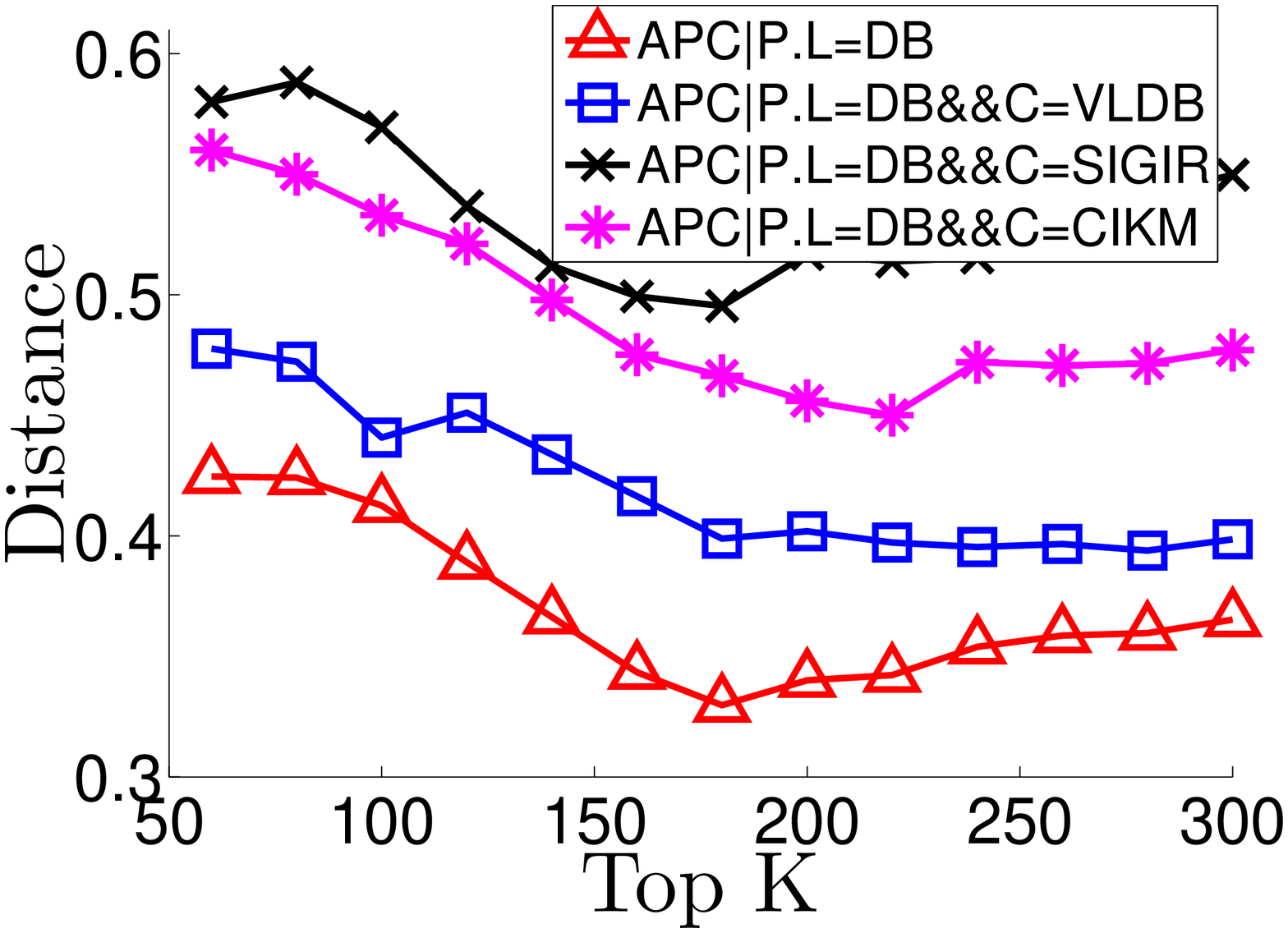}
\end{minipage}
}%
\subfigure[IR field]{
\begin{minipage}[t]{0.3\textwidth}
  \includegraphics[width=4cm]{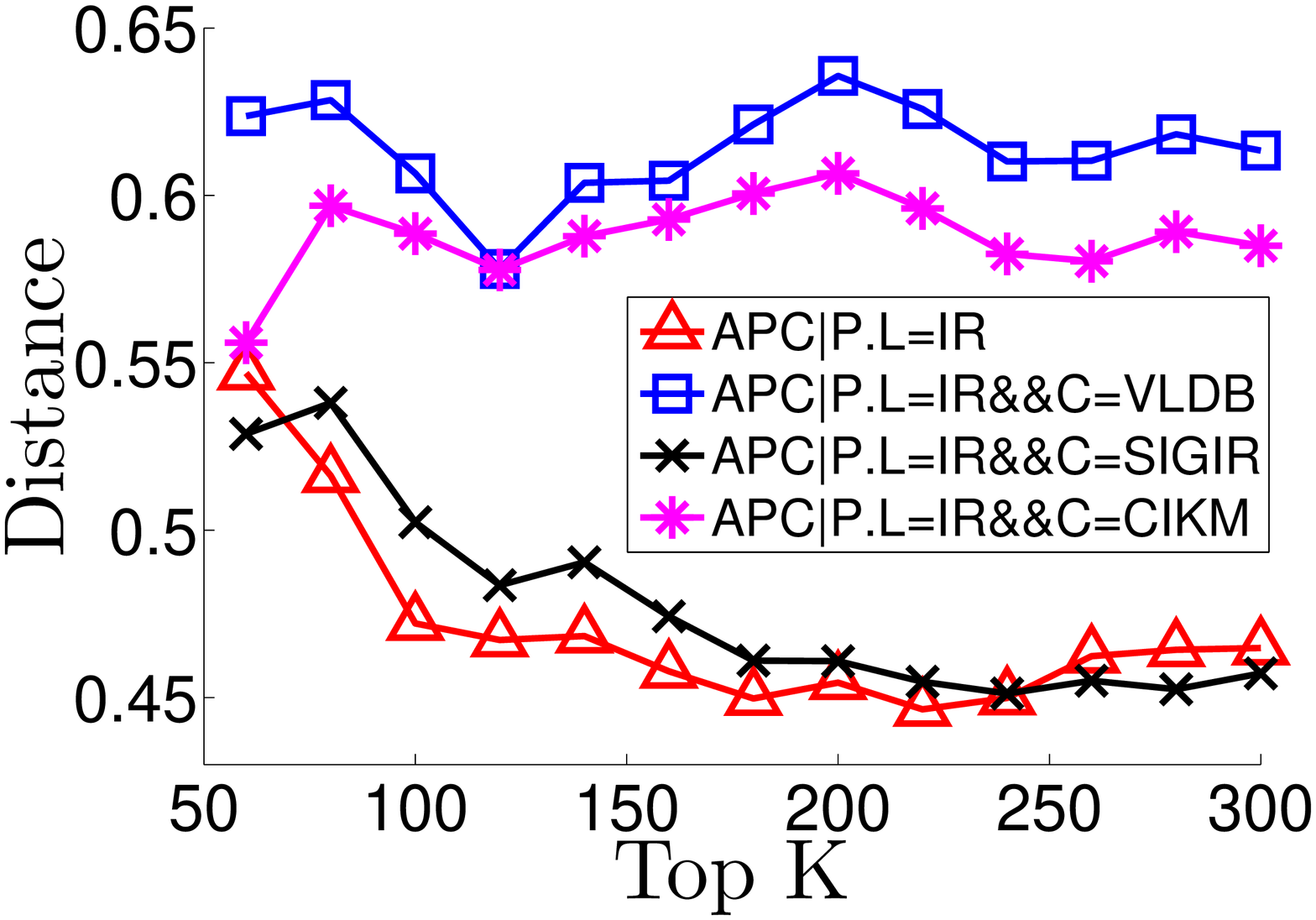}
\end{minipage}
}
\caption{\small The rank accuracy of HRank with different constrained meta paths on DBLP dataset.}\label{fig:fft}
\end{figure}

%\vspace{-10pt}
\begin{figure}[htbp]
\centering
\small
\subfigure[Authors]{\label{fig:fft:a}
\begin{minipage}[t]{0.23\textwidth}
  \includegraphics[width=4.5cm]{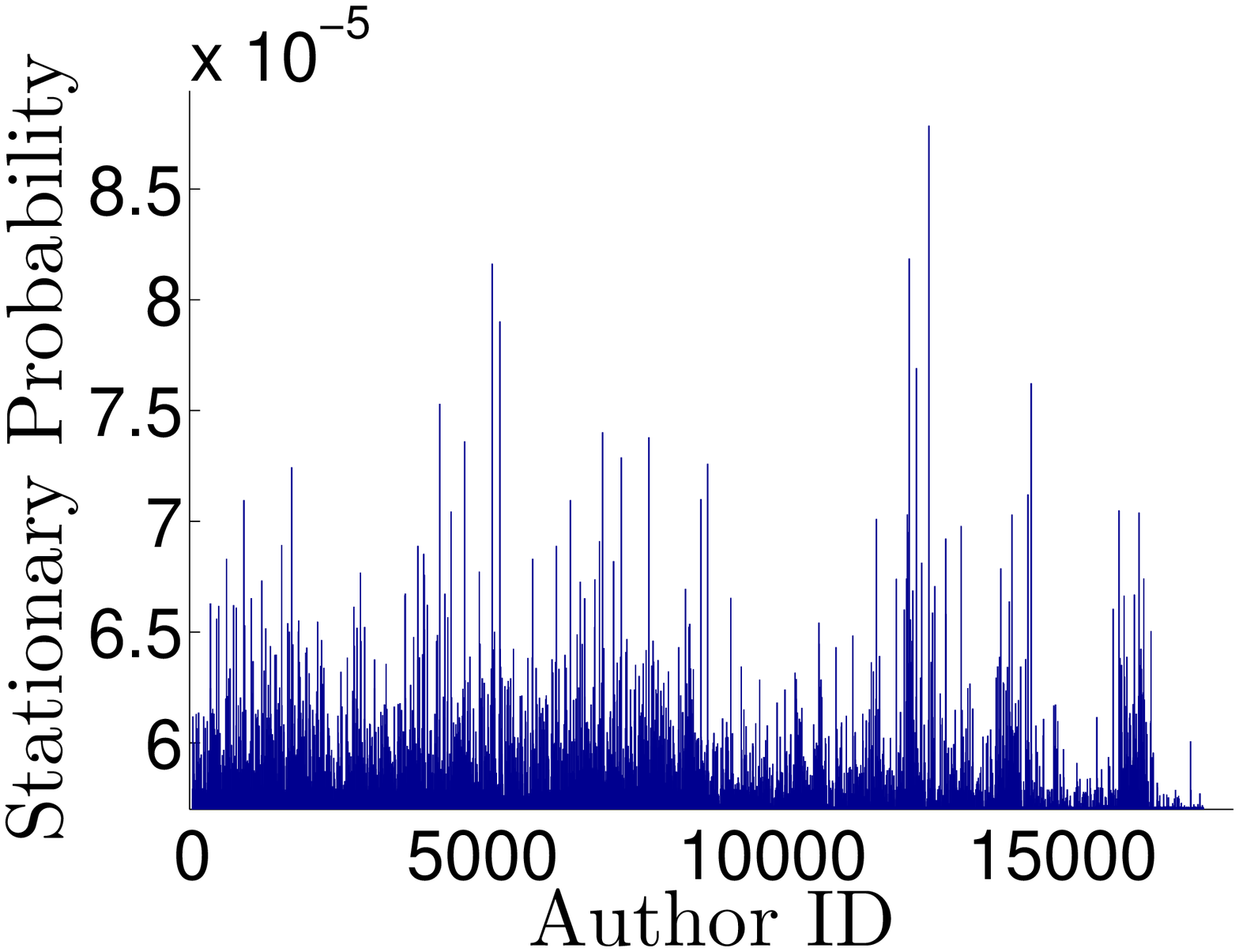}
\end{minipage}%
}%
\subfigure[Paths]{
\begin{minipage}[t]{0.3\textwidth}
  \includegraphics[width=4.5cm]{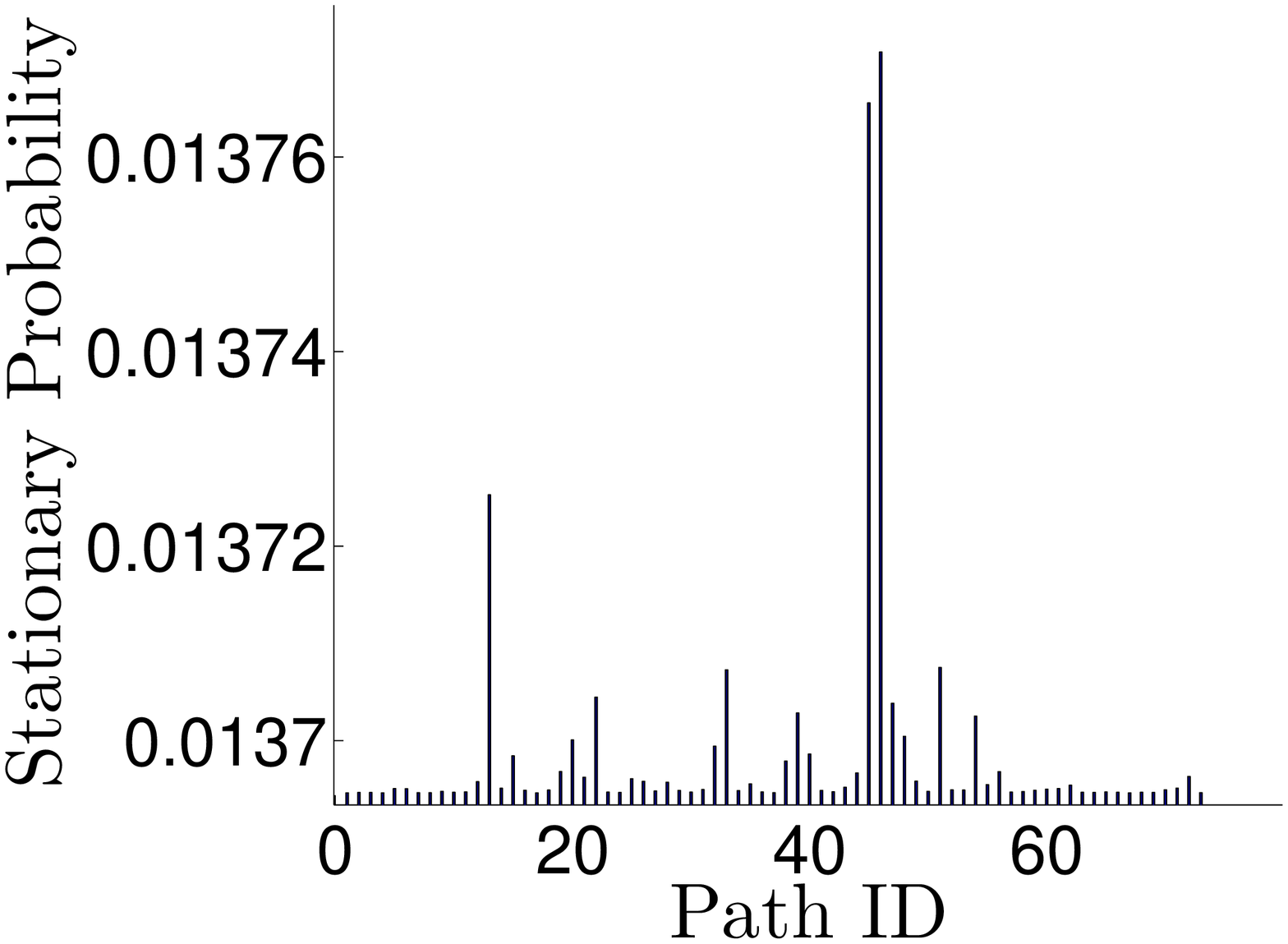}
\end{minipage}
}%\vspace{-5pt}
\caption{\small The stationary probability distributions of authors and constrained meta paths.}\label{fig:fft}
\end{figure}
%\vspace{-10pt}

\begin{table}[htbp]
\caption{\small Top 10 authors and constrained meta paths (note that only the constraint ($L_j$) of the paths ($APA|P.L = L_j, j = 1\ldots 73$) are shown in the third column of the table).}
\begin{center}
\resizebox{.47\textwidth}{!}{
\begin{tabular}{|c|c|c|}
\hline
Rank & \multicolumn{1}{c|}{Authors} & \multicolumn{1}{c|}{Constrained meta paths} \\\hline
1& {Jiawei Han}& {H.3 (Information Storage and Retrieval)}   \\
2& {Philip Yu}& {H.2 (Database Management)} \\
3& {Christos Faloutsos}& {C.2 (Computer-Communication Networks)}\\
4& {Ravi Kumar}& {I.2 (Artificial Intelligence)}  \\
5& {Wei-Ying Ma}& {F.2 (Analysis of Algorithms and Problem Complexity)}  \\
6& {Zheng Chen}& {D.4 (Operating Systems)}  \\
7& {Hector Garcia-Molina}& {H.4 (Information Systems Applications)}  \\
8& {Hans-Peter Kriegel}& {G.2 (Discrete Mathematics)}  \\
9& {Gerhard Weikum}& {I.5 (Pattern Recognition)} \\
10& {D. R. Karger}& {H.5 (Information Interfaces and Presentation)} \\\hline
\end{tabular}
}
\end{center}
\end{table}

\subsection{Co-Ranking of Objects and Paths}
\subsubsection{Experiment Study on Co-Ranking on Symmetric Constrained Meta Paths}

In this experiment, we will validate the effectiveness of HRank-CO to rank objects and symmetric constrained meta paths simultaneously. The experiment is done on ACM dataset. First we construct a (2, 1)th order tensor $X$ based on 73 constrained meta paths (i.e., $APA|P.L = L_j, j = 1\cdots 73$). When the $i$th and the $k$th authors co-publish a paper together, of which the label is the $j$th label (i.e., ACM categories), we add one to the entries $x_{i,j,k}$ and $x_{k,j,i}$ of $X$. In this case, $X$ is symmetric with respect to the index $j$. By considering all the publications, $x_{i,j,k}$ (or $x_{k,j,i}$ ) refers to the number of collaborations by the $i$th and the $k$th author under the $j$th paper label. In addition, we do not consider any self-collaboration, i.e., $x_{i,j,i}$ = 0 for all $1 \leq i \leq 17431$ and $1 \leq j \leq 73$. The size of $X$ is $17431 \times 73 \times 17431$ where there are 91520 nonzero entries in $X$. The percentage of nonzero entries is $4.126 \times 10^{-4}\%$. In this dataset, we will evaluate the importance of authors through the co-author relations, meanwhile we will analyze the importance of paths (i.e., which paths have the most contributions to the importance of authors).

\begin{table*}[htbp]
\caption{ The number that the top ten authors collaborate with others via the top ten constrained meta paths (note that only the constraint ($L_j$) of the paths ($APA|P.L = L_j, j = 1\ldots 73$) are shown in the first row of the table).}
\begin{center}
\resizebox{.7\textwidth}{!}{
\begin{tabular}{|c|cccccccccc|}
\hline
Ranked A/$CP$ & \multicolumn{1}{c}{1 (H.3)} & \multicolumn{1}{c}{2 (H.2)} & \multicolumn{1}{c}{3 (C.2)} & \multicolumn{1}{c}{4 (I.2)} & \multicolumn{1}{c}{5 (F.2)} & \multicolumn{1}{c}{6 (D.4)} & \multicolumn{1}{c}{7 (H.4)} & \multicolumn{1}{c}{8 (G.2)} & \multicolumn{1}{c}{9 (I.5)} & \multicolumn{1}{c|}{10 (H.5)} \\\hline
1 (Jiawei Han)& 51& 176& 0& 0& 0& 0& 9& 2& 2& 0 \\
2 (Philip Yu)& 51& 94& 0& 0& 9& 0& 3& 0& 13& 0\\
3 (C. Faloutsos)& 17& 107& 0& 5& 9& 0& 3& 4& 2& 0 \\
4 (Ravi Kumar)& 73& 27& 0& 3& 13& 0& 18& 5& 0& 0 \\
5 (Wei-Ying Ma)& 132& 26& 0& 9& 0& 0& 2& 0& 30& 10 \\
6 (Zheng Chen)& 172& 9& 0& 9& 0& 0& 22& 0& 38& 9 \\
7 (H. Garcia-Molina)& 23& 65& 3& 0& 0& 0& 1& 0& 0&	4 \\
8 (H. Kriegel)& 19& 28& 5& 0& 0& 0& 6& 0& 7&	4 \\
9 (G. Weikum)& 82& 14&	0& 4& 0& 0& 8& 0& 4& 0 \\
10 (D. R. Karger)& 11& 5&	13& 0& 7& 4& 1& 7& 0& 7 \\\hline
\end{tabular}
}
\end{center}
\end{table*}

Figure 8 shows the stationary probability distributions of authors and paths. It is obvious that some authors and paths have higher stationary probability, which implies these authors and paths are more important than others. Table 4 shows the top ten authors (left) and paths (right) based on their HRank values. We can find that the top ten authors all are influential researchers in the DM/IR fields, which conforms to our common senses. Similarly, the most important paths are related to DM/IR fields, such as $APA|P.L =``H.3"$ (Information Storage and Retrieval) and $APA|P.L =``H.2"$ (Database Management). Although the conferences in ACM dataset are from multiple fields, such as DM/DB (e.g., KDD, SIGMOD) and computation theory (e.g., SODA, STOC), there are more papers from the DM/DB fields, which makes the authors and paths in the DM/DB fields ranked higher. We can also find that the influence of authors and paths can be promoted by each other. The reputation of Jiawei Han and Philip Yu come from their productive papers in the influential fields (e.g., H.3 and H.2). In order to observe this point more clearly, we show the number of co-authors of the top ten authors based on the top ten paths in Table 5. We can observe that there are more collaborations for top authors in the influential fields. For example, although Zheng Chen (rank 6) has more number of co-authors than Jiawei Han (rank 1), the collaborations of Jiawei Han focus on ranked higher fields (i.e., H.3 and H.2), so Jiawei Han has higher HRank score. Similarly, the top paths contain many collaborations of influential authors.

\subsubsection{Experiment Study on Co-Ranking on Asymmetric Constrained Meta Paths}

The experiments on the Movie dataset aim to show the effectiveness of HRank-CO to rank heterogeneous objects and asymmetric constrained meta paths simultaneously. In this case, we construct a 3rd order tensor $X$ based on the constrained meta paths $AMD|M.T$. That is, the tensor represents the actor-director collaboration relations on different types of movies. When the $i$th actor and the $k$th director cooperate in a movie of the $j$th type, we add one to the entries $x_{i,j,k}$ of $X$. By considering all the cooperations, $x_{i,j,k}$ refers to the number of collaborations by the $i$th actor and the $k$th director under the $j$th type of movie. The size of $X$ is $5324 \times 112 \times 551$ and there are 36529 nonzero entries in $X$. The percentage of nonzero entries is $7.827 \times 10^{-4}\%$.

\begin{table}[!t]
\caption{\small Top 10 actors, directors and meta paths on IMDB dataset (note that only the constraint ($T_j$) of the paths ($AMD|M.T = T_j, j = 1\ldots 1591$) are shown in the fourth column).}
\begin{center}
\resizebox{.47\textwidth}{!}{
\begin{tabular}{|c|c|c|c|}
\hline
Rank & \multicolumn{1}{c|}{Actor} & \multicolumn{1}{c|}{Director} & \multicolumn{1}{c|}{Conditional meta path}\\\hline
1& {Eddie Murphy}& {Tim Burton}& {Comedy}   \\
2& {Harrison Ford}& {Zack Snyder}& {Drama}  \\
3& {Bruce Willis}& {Marc Forster}& {Thriller}  \\
4& {Drew Barrymore}& {David Fincher}& {Action}  \\
5& {Nicole Kidman}& {Michael Bay}& {Adventure}  \\
6& {Nicolas Cage}& {Ridley Scott}& {Romance}  \\
7& {Hugh Jackman}& {Richard Donner}& {Crime}  \\
8& {Robert De Niro}& {Steven Spielberg}& {Sci-Fi}  \\
9& {Brad Pitt}& {Robert Zemeckis}& {Animation}  \\
10& {Christopher Walken}& {Stephen Sommers}& {Fantasy} \\\hline
\end{tabular}
}

\end{center}
\end{table}

Table 6 shows the top ten actors, directors and constrained meta paths (i.e., movie type). We observe the mutual enhancements of the importance of objects and meta paths again. Basically, the results comply with our common senses. The top ten actors are well known, such as Eddie Murphy, Harrison Ford. Similarly, these directors are also famous in filmdom due to their works. These movie types obtained are the most popular movie subjects as well. In addition, we can observe the mutual effect of objects and paths one more time. As we know, Eddie Murphy and Drew Barrymore (rank 1, 4 in actors) are famous comedy and drama (rank 1, 2 in paths) actors. Harrison Ford and Bruce Willis (rank 2,3 in actors) are popular thrill and action (rank 3,4 in paths) actors. These higher ranked directors also prefer those popular movie subjects.

\begin{figure*}[t]
    \begin{center}
    \subfigure[Running time on ${(APA)}^l$]
    {\includegraphics[scale=0.18]{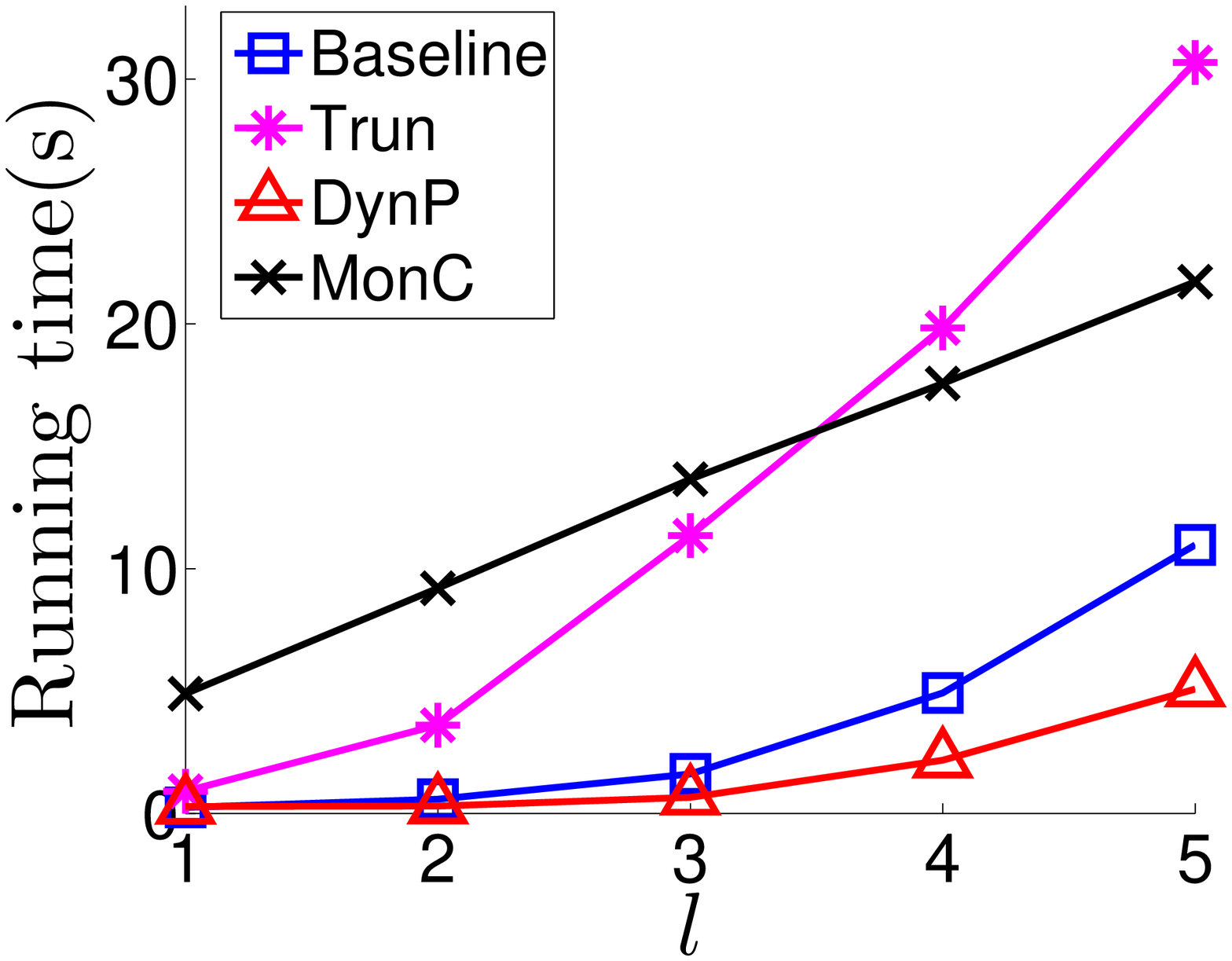}}
    \subfigure[Running time on ${(APCPA)}^l$]
    {\includegraphics[scale=0.18]{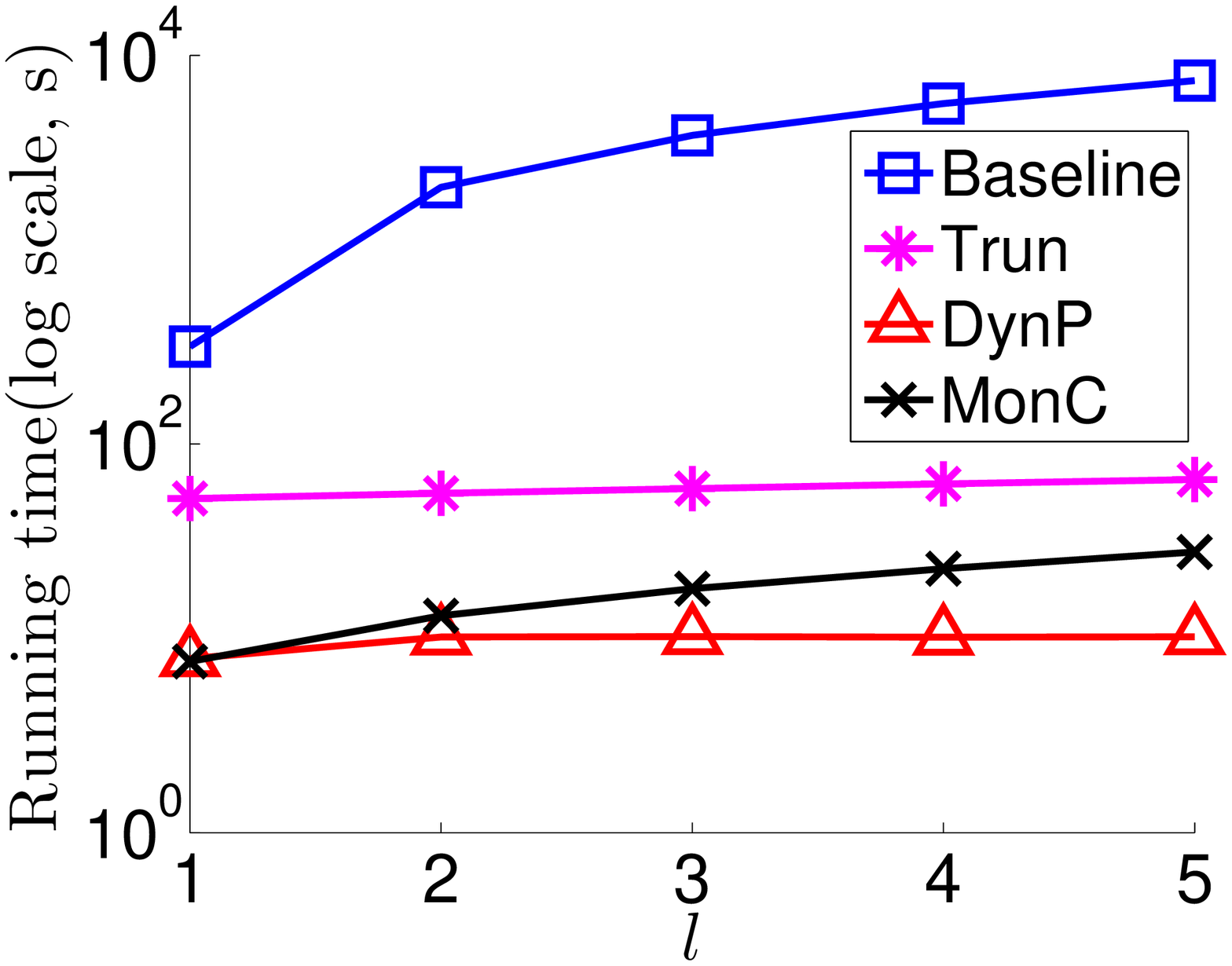}}
    \subfigure[Accuracy on ${(APA)}^l$]
    {\includegraphics[scale=0.18]{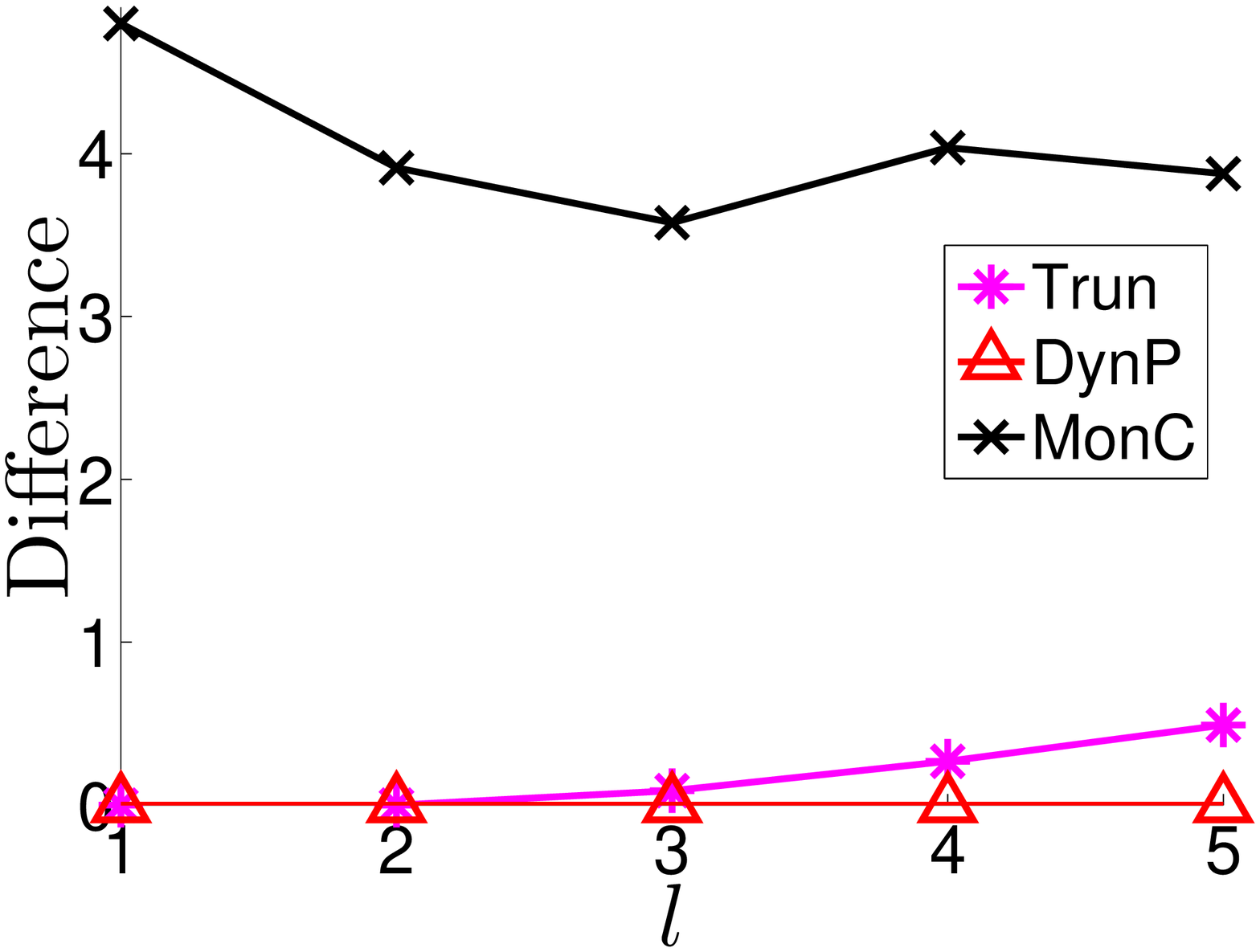}}
    \subfigure[Accuracy on ${(APCPA)}^l$]
    {\includegraphics[scale=0.18]{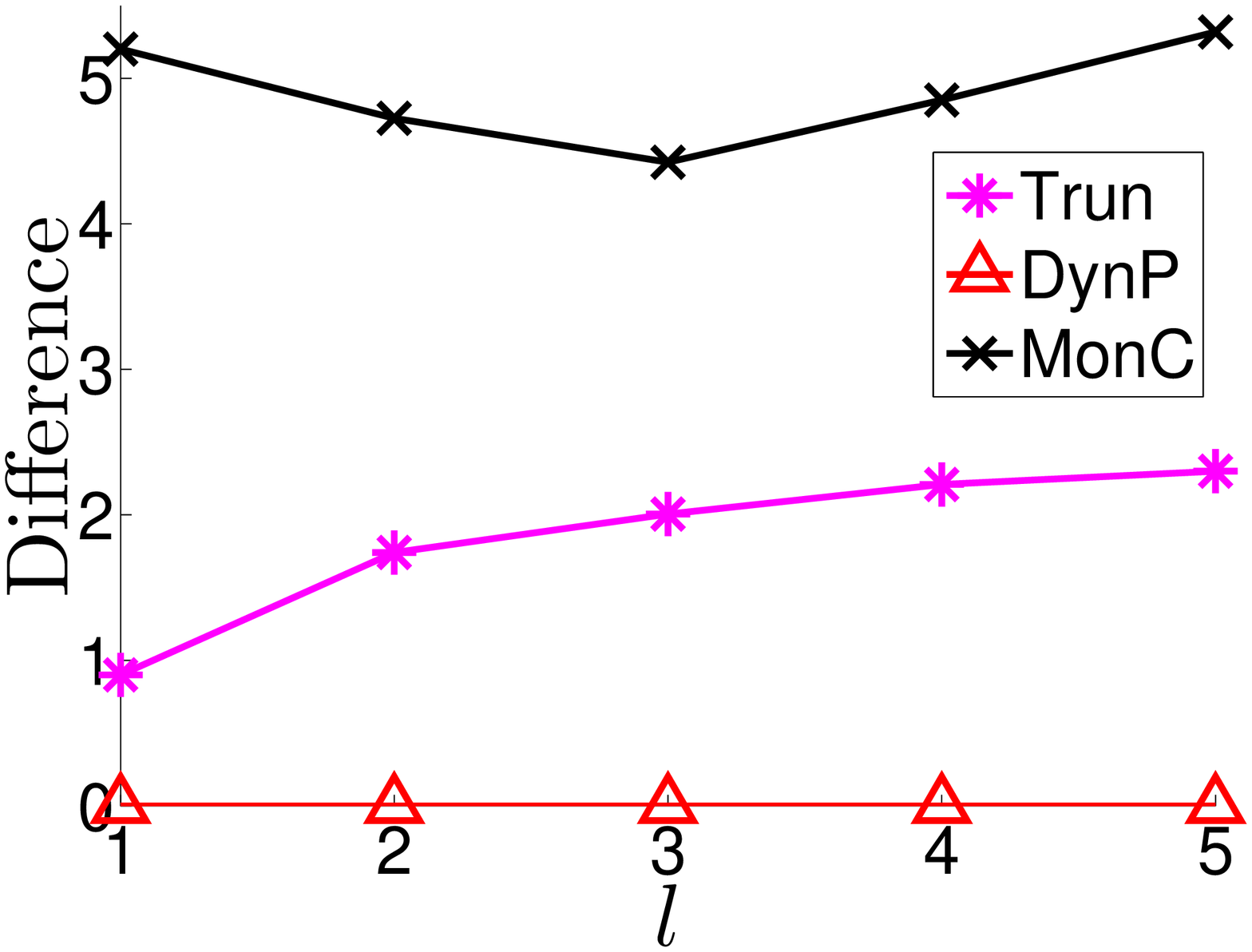}}
    \end{center}
    \caption{Running time and accuracy of matrix multiplication based on different fast computation strategies and paths.}\label{PartRunTime}
\end{figure*}

\subsection{Fast Computation Experiments}
Based on the DBLP dataset, we select two meta paths
with varying length ($l$): ${(APA)}^l$ and ${(APCPA)}^l$,
where $l$ means times of path repetition ranging from 1 to 5. We record
the running time of matrix multiplication based on these paths with different fast computation strategies. The direct matrix multiplication is baseline. Meanwhile, we calculate the differences of results obtained by baseline and fast computation strategies (i.e., $F$ norm of differences of two matices). These differences can be considered as the accuracy measure of fast computation strategies (the smaller the better). We set the parameters in the Trun strategy as follows: $W$ is 200, $\beta$ is 0.5, and $\gamma$ is 0.02. The number of walkers (i.e., $K$) in the MonC strategy is 500. All experiments are done on machines with Intel Xeon 8-Core CPUs of 2.13 GHz and 64 GB RAM.

Figure 9 shows the running time and accuracy of three strategies on different paths. From Figure 9(a) and (b), we can find that the DynP is an effective strategy to fasten matrix multiplication on both paths, while the Trun and MonC strategies only speed up matrix multiplication on the path ${(APCPA)}^l$. During the matrix multiplication along ${(APA)}^l$, the matrix is always sparse, so the baseline itself is very fast. In this condition, the Trun and MonC strategies do not work. For the path ${(APCPA)}^l$, the multiplication matrix becomes dense due to the low dimension of $C$ (\# of conferences is 20), so its running time increases greatly. In this condition, the Trun and MonC are also effective strategies to fasten matrix multiplication. Then, we observe their accuracy from Figure 9 (c) and (d). As an information-lossless strategy, the DynP's results are the same as with baseline. The MonC strategy has the lowest accuracy. To sum up, the DynP can effectively accelerate matrix multiplication without loss on accuracy, while the Trun and MonC strategies also help to speed up the multiplication of dense matrices.

In Section 4.4, we have pointed out that the time complexity of the rank computation in HRank-CO is $O(trn^2)$. However, for real applications, the relation matrices are very sparse, so the real time complexity is linear to the number of links. This point is confirmed by the following experiments. We create three different scales of tensors (size $n\times r\times n$). We record the running time for rank computation on different link densities. The results are shown in Figure 10. It is clear that the running time slowly and near linearly increases with the increment of link density. Moreover, the longer running time is needed for larger-scale tensor.

%\vspace{-10pt}
\begin{figure}[htbp]

\center
  \includegraphics[width=4.2cm]{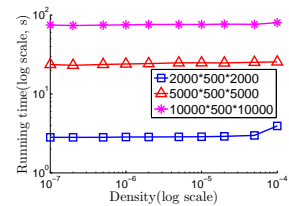}

\caption{\small The running time for rank computation on different link densities.}\label{fig:fft}
\end{figure}

\subsection{Convergence Experiments}

In Figure 11, we show the convergence of HRank on the previous experiments. The results illustrate that the three versions of HRank all quickly converge after no more than 20 iterations. In addition, we can also observe that HRank has different convergence speed in these three conditions. For symmetric meta paths, the HRank-SY almost converges on 6 iterations (see Figure 11(a)). However, HRank-CO for co-ranking converges on 16 iterations (see Figure 11(c)). We think it is reasonable, since it is more difficult to converge for more objects in the HRank-CO.

\begin{figure}[htbp]
    \begin{center}
    \subfigure[HRank-SY]
    {\includegraphics[scale=0.11]{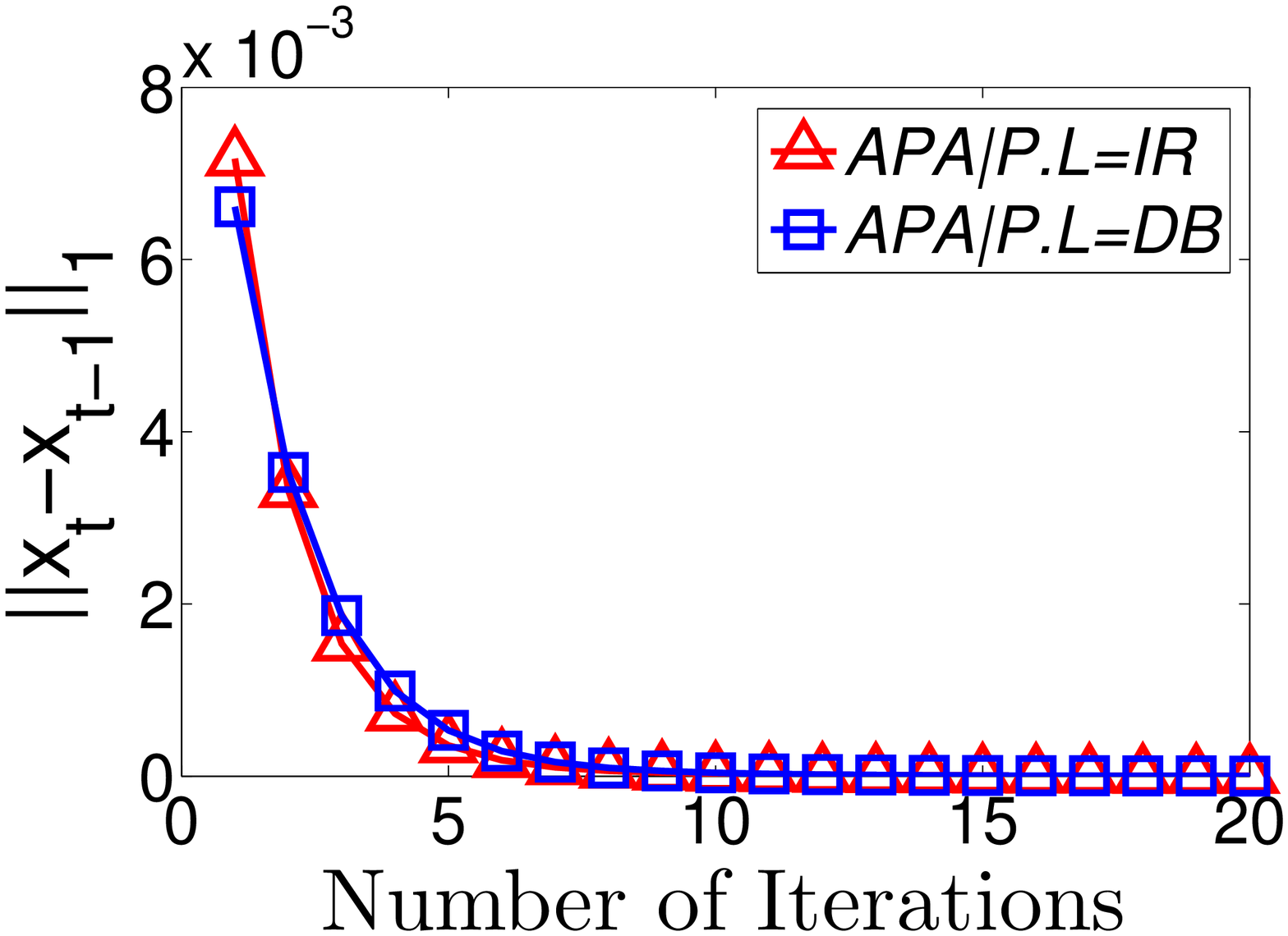}}
    \subfigure[HRank-AS]
    {\includegraphics[scale=0.11]{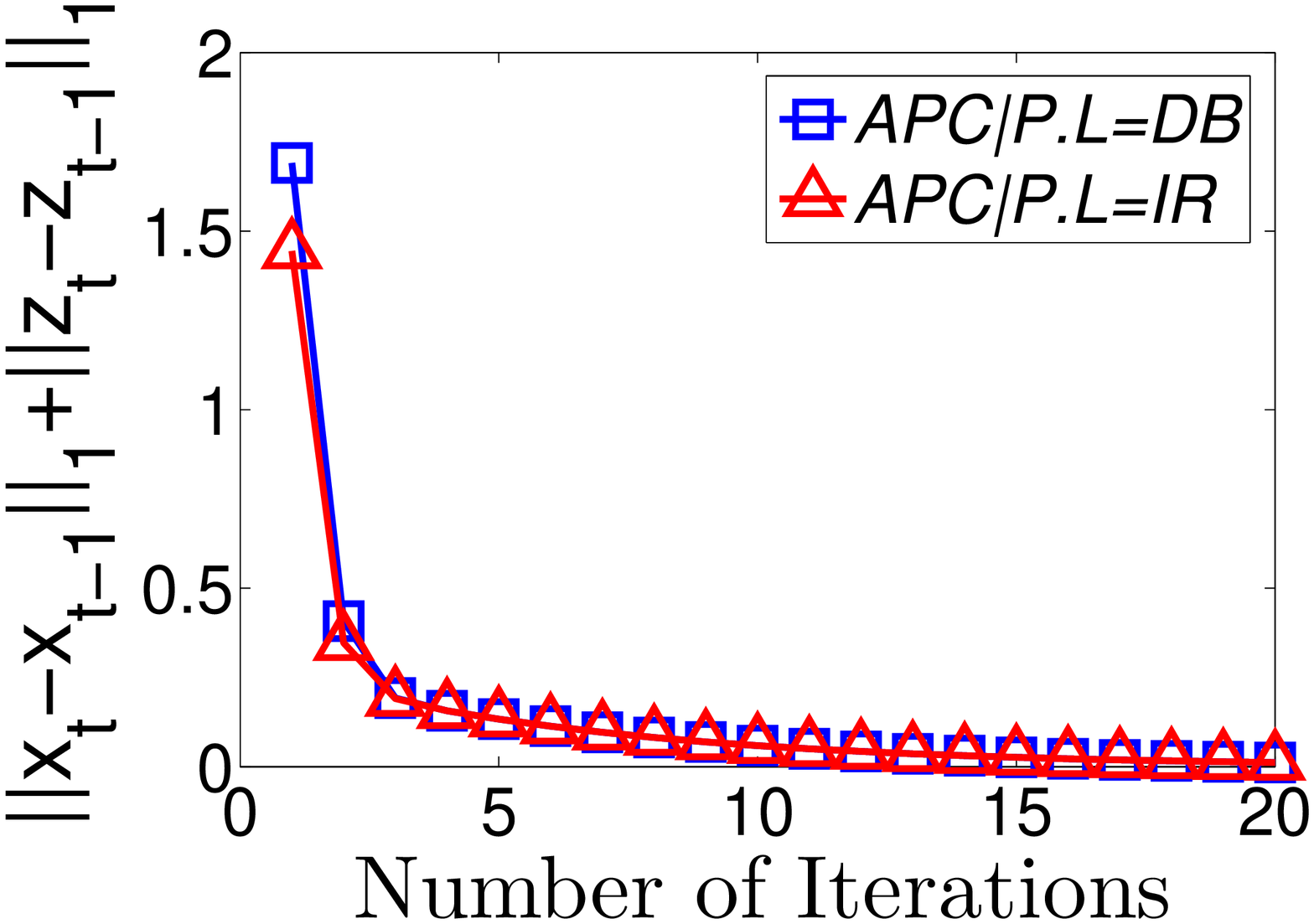}}
    \subfigure[HRank-CO]
    {\includegraphics[scale=0.11]{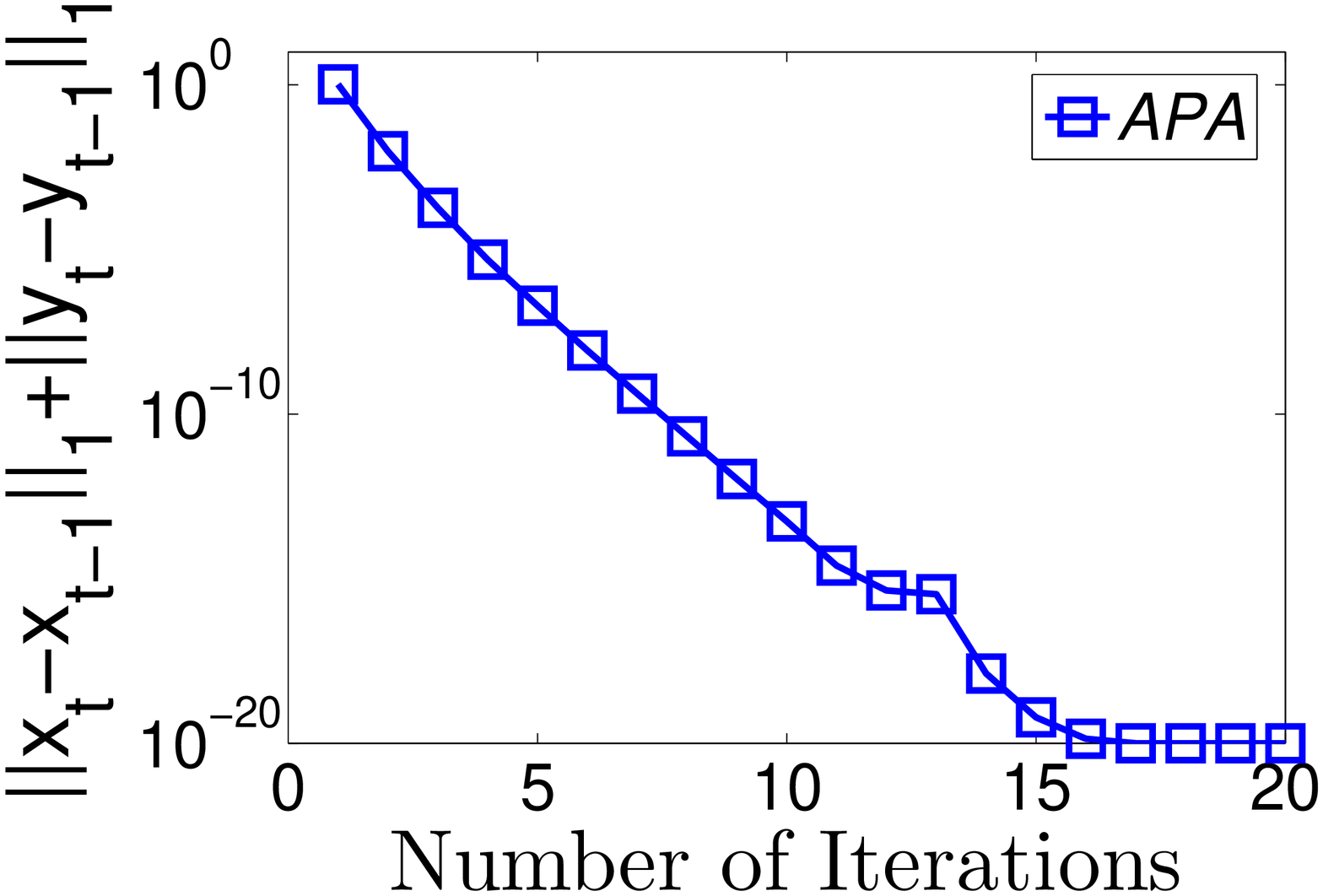}}
    \end{center}
\caption{\small The difference between two successive calculated probability vectors against iterations based on the three versions of HRank.}\label{fig:fft}
\end{figure}

\section{Conclusions}
In this paper, we first study the ranking problem in heterogeneous information network and propose the HRank framework, which is a path based random walk method. In this framework, we introduce the constrained meta path concept to capture the more subtle and refined semantics contained in HIN. In addition, we further put forward a method to co-rank the paths and objects, since the paths effect the importance of objects. Extensive experiments validate the effectiveness and efficiency of HRank on three real datasets.

%\end{document}  % This is where a 'short' article might terminate

%ACKNOWLEDGMENTS are optional
%\section{Acknowledgments}
%This section is optional; it is a location for you
%to acknowledge grants, funding, editing assistance and
%what have you.  In the present case, for example, the
%authors would like to thank Gerald Murray of ACM for
%his help in codifying this \textit{Author's Guide}
%and the \textbf{.cls} and \textbf{.tex} files that it describes.

\bibliographystyle{abbrv}
\bibliography{References}

\begin{thebibliography}{10}

\bibitem{SC07}
S.~Chakrabarti.
\newblock Dynamic personalized pagerank in entity-relation graphs.
\newblock In {\em WWW}, pages 571--580, 2007.

\bibitem{DLK09}
H.~Deng, M.~R. Lyu, and I.~King.
\newblock A generalized co-hits algorithm and its application to bipartite
  graphs.
\newblock In {\em KDD}, pages 239--248, 2009.

\bibitem{Han09}
J.~Han.
\newblock Mining heterogeneous information networks by exploring the power of
  links.
\newblock In {\em Discovery Science}, pages 13--30, 2009.

\bibitem{JW02}
G.~Jeh and J.~Widom.
\newblock Simrank: a measure of structural-context similarity.
\newblock In {\em KDD}, pages 538--543, 2002.

\bibitem{JSDHG10}
M.~Ji, Y.~Sun, M.~Danilevsky, J.~Han, and J.~Gao.
\newblock Graph regularized transductive classification on heterogeneous
  information networks.
\newblock In {\em ECML/PKDD}, pages 570--586, 2010.

\bibitem{JLH11}
R.~Jin, V.~E. Lee, and H.~Hong.
\newblock Axiomatic ranking of network role similarity.
\newblock In {\em KDD}, pages 922--930, 2011.

\bibitem{K99}
J.~M. Kleinberg.
\newblock Authoritative sources in a hyperlinked environment.
\newblock In {\em SODA}, pages 668--677, 1999.

\bibitem{KYDW}
X.~Kong, P.~S. Yu, Y.~Ding, and D.~J. Wild.
\newblock Meta path-based collective classification in heterogeneous
  information networks.
\newblock In {\em CIKM}, pages 1567--1571, 2012.

\bibitem{LC10a}
N.~Lao and W.~Cohen.
\newblock Fast query execution for retrieval models based on path constrained
  random walks.
\newblock In {\em KDD}, pages 881--888, 2010.

\bibitem{LNY12}
X.~Li, M.~K. Ng, and Y.~Ye.
\newblock Har: Hub, authority and relevance scores in multi-relational data for
  query search.
\newblock In {\em SDM}, pages 141--152, 2012.

\bibitem{NLY11}
M.~K. NG, X.~Li, and Y.~Ye.
\newblock Multirank: Co-ranking for objects and relations in multi-relational
  data.
\newblock In {\em KDD}, pages 1217--1225, 2011.

\bibitem{NZWM05}
Z.~Nie, Y.~Zhang, J.~Wen, and W.~Ma.
\newblock Object-level ranking: bringing order to web objects.
\newblock In {\em WWW}, pages 422--433, 2005.

\bibitem{PBMW98}
L.~Page, S.~Brin, R.~Motwani, and T.~Winograd.
\newblock The pagerank citation ranking: Bringing order to the web.
\newblock Technical report, Stanford University Database Group, 1998.

\bibitem{SKYX12}
C.~Shi, X.~Kong, P.~S. Yu, S.~Xie, and B.~Wu.
\newblock Relevance search in heterogeneous networks.
\newblock In {\em EDBT}, pages 180--191, 2012.

\bibitem{SZKYLW12}
C.~Shi, C.~Zhou, X.~Kong, P.~S. Yu, G.~Liu, and B.~Wang.
\newblock Heterecom: A semantic-based recommendation system in heterogeneous
  networks.
\newblock In {\em KDD}, pages 1552--1555, 2012.

\bibitem{SHYYW11}
Y.~Sun, J.~Han, X.~Yan, P.~Yu, and T.~Wu.
\newblock Pathsim: Meta path-based top-k similarity search in heterogeneous
  information networks.
\newblock In {\em VLDB}, pages 992--1003, 2011.

\bibitem{SHZYCW09}
Y.~Sun, J.~Han, P.~Zhao, Z.~Yin, H.~Cheng, and T.~Wu.
\newblock {RankClus}: integrating clustering with ranking for heterogeneous
  information network analysis.
\newblock In {\em EDBT}, pages 565--576, 2009.

\bibitem{SNHYYY12}
Y.~Sun, B.~Norick, J.~Han, X.~Yan, P.~S. Yu, and X.~Yu.
\newblock Integrating meta-path selection with user-guided object clustering in
  heterogeneous information networks.
\newblock In {\em KDD}, pages 1348--1356, 2012.

\bibitem{SYH09}
Y.~Sun, Y.~Yu, and J.~Han.
\newblock Ranking-based clustering of heterogeneous information networks with
  star network schema.
\newblock In {\em KDD}, pages 797--806, 2009.

\bibitem{YSNMH12}
X.~Yu, Y.~Sun, B.~Norick, T.~Mao, and J.~Han.
\newblock User guided entity similarity search using meta-path selection in
  heterogeneous information networks.
\newblock In {\em CIKM}, pages 2025--2029, 2012.

\bibitem{ZOZG07}
D.~Zhou, S.~A. Orshanskiy, H.~Zha, and C.~L. Giles.
\newblock Co-ranking authors and documents in a heterogeneous network.
\newblock In {\em ICDM}, pages 739--744, 2007.

\end{thebibliography}
% ACM needs 'a single self-contained file'!
%

\balancecolumns
% That's all folks!
\end{document}